\documentstyle[12pt,epsf,a4]{article}
\newcommand{\be}{\begin{equation}}
\newcommand{\ee}{\end{equation}}
\newcommand{\ba}{\begin{eqnarray}}
\newcommand{\ea}{\end{eqnarray}}
\newcommand{\bi}{\begin{itemize}}
\newcommand{\ei}{\end{itemize}}

\def\lsi{\raise0.3ex\hbox{$<$\kern-0.75em\raise-1.1ex\hbox{$\sim$}}}
\def\gsi{\raise0.3ex\hbox{$>$\kern-0.75em\raise-1.1ex\hbox{$\sim$}}}

\makeatletter \@addtoreset{equation}{section} \makeatother

\begin{document}

\begin{titlepage}
\begin{flushright}
LTH-420\\
hep-lat/9802015\\
February 10, 1998\\
\end{flushright}
\begin{centering}
\vfill

{\bf Maximal variance reduction for stochastic  propagators with 
applications to the static quark spectrum}
\vspace{0.8cm}

{UKQCD Collaboration}

\vspace{0.4cm}

C. Michael\footnote{cmi@liv.ac.uk} and
J. Peisa\footnote{peisa@amtp.liv.ac.uk} \\

\vspace{0.3cm}
{\em Department of Mathematical Sciences, 
University of Liverpool, \\ Liverpool L69 3BX, UK}

\vspace{0.7cm}
{\bf Abstract}

\end{centering}

\vspace{0.3cm}\noindent
 We study a new method --  maximal variance reduction -- for reducing
the  variance of stochastic estimators for  quark propagators. We find
that while this method is comparable to usual iterative inversion for
light-light mesons, a considerable improvement is achieved for systems
containing at least one infinitely heavy quark. Such systems are needed
for heavy quark effective theory. As an illustration of the
effectiveness  of the method we  present results for the masses of the
ground state  and excited states of $\bar{Q}q$  mesons and $\bar{Q}qq$
baryons. We compare  these results with the experimental spectra
involving $b$ quarks.
 \vfill
\noindent

\end{titlepage}

\section{Introduction}
 When computing hadron-hadron correlators, one needs quark propagators
from a given source to a sink. Optimally one would like to calculate the
propagators required from all sites to all sites, and thus use all
information available from the finite number of gauge samples. In
practice this is seldom possible as one has to invert the Wilson-Dirac
fermion matrix to obtain the propagators. Using conventional iterative
methods, one obtains propagators from one source to all given sinks; to
calculate and store such propagators from all sources is virtually
impossible with current computing resources. Furthermore there is no
known way to stop iterating before one reaches the machine precision
without introducing bias. Therefore one obtains extremely accurate
propagators from few sources. The propagators are so accurate that the 
variance coming from the limited sample of gauge configurations
dominates the results  totally. Clearly a lot of time is wasted on
calculating the propagators to such precision, when the variance from
one gauge configuration to another is several orders of magnitude
larger. 

One possibility is to calculate also the propagators by Monte Carlo
methods~\cite{dfmp,tsukuba,pm}. This allows one to store the propagators
from everywhere to everywhere in a sensible amount of storage space and
also avoids the unnecessary calculation of the propagators to machine
precision.

It is easy to express the inverse of a positive definite matrix $A$ in a
form suitable for Monte Carlo integration: one just takes a Gaussian
integral
 \be 
 A_{ij}^{-1} = {1\over Z} \int D\phi\, \phi_j^* \phi_i \exp(-{1 \over 2}
\phi^* A \phi), \label{stokinv}
 \ee
 which then can be treated exactly as a free scalar field on the
lattice. If the matrix $A$ is local, it is  easy to implement efficient
Monte Carlo update techniques  for the scalar field $\phi$,  allowing
one to calculate the required inverse of $A$. Thus, for a given  gauge
field, one would obtain $N$ independent samples of the $\phi$ fields  by
Monte Carlo and so evaluate the stochastic estimate of the required
element of the inverse of $A$  by an average over these $N$ samples: $
A_{ij}= \langle \phi_j^* \phi_i \rangle$. By storing these $N$ samples
of $\phi$, one would then be able to evaluate propagators from any  site
to any site.

This is not directly applicable to the Wilson-Dirac fermion matrix
$Q=1-\kappa M$, because $Q$ is not positive definite for those values of
hopping parameter $\kappa$ that one is usually  interested in. To obtain
the propagators by the above method, one has to work with $A=Q^\dagger
Q$, which is guaranteed to be positive definite. As $Q$ contains only
nearest neighbour interactions, $A$ is  still local -- it contains at
most next-to-nearest neighbour interactions, and an effective updating
scheme can be implemented. Of course to recover the inverse of $Q$
instead of $A^{-1}$ one should modify eq.~(\ref{stokinv}) to  
 \be 
G_{ji} = Q_{ji}^{-1} = \left< (Q_{ik}\phi_k)^* \phi_j \right>,
\label{propagator} 
 \ee 
 which can then be used to calculate the propagators one needs for
hadronic observables.

In practice a direct application of eq.~(\ref{propagator}) has a serious
drawback when used in realistic lattice QCD calculations. Because the
$\phi$ fields have a variance of order one coming  from the gaussian
distribution which determines them,  the standard deviation on the
estimate of the propagator will be of order $N_s^{-1/2}$  for $N_s$
samples of $\phi$ fields.  Usually one is interested in the  large $T$
behaviour of the correlators of  hadronic observables. These correlators
decay exponentially, and therefore the signal is exponentially small
(like $\exp(-mT)$ where $m$ is the hadron mass) in the regime of
interest. As the variance of the propagators calculated from
eq.~(\ref{propagator}) is the same no matter how far in time they
extend, it would be necessary to use impractical amounts of computer
time to increase the number of samples $N_s$ sufficiently to obtain a
reasonable  signal to noise ratio at large $T$.

In this paper we will discuss stochastic  methods to calculate
propagators. We will compare several suggestions to avoid the problems
described above, and show that it is possible to construct propagators
from scalar fields  that have their variance maximally reduced. We also
discuss the use of improved fermionic actions with stochastic
estimators.

To test different methods, we focus our attention to systems which
contain one infinitely heavy quark as  obtained in leading
order heavy quark  effective theory\cite{hqet}. Such a study is
appropriate in particular to the B meson and its excited states which 
are made of one heavy quark and one light quark. These systems are
particularly problematic for conventional methods of evaluating light
quark propagators, because, using  one source for the light quark
propagator, there will be only one measurement of the hadronic
correlator to time $T$ per gauge configuration when the heavy quark is
treated as static.  Therefore it seems that one would benefit
hugely from having propagators from all sources available. This benefit
will then  help to offset the extra noise coming from having only a
stochastic estimate.

 For mesonic correlations, the signal is linear in the light quark
propagator and so no problems arise with biases or correlations among
stochastic samples. A  more careful analysis is needed for observables
involving more than one light quark propagator.  Here we study one
example in detail: the baryonic system made of one static quark and  two
light quarks.

We also study the feasibility of applying our method of choice --
maximal variance reduction -- to systems where all the quarks
are propagating. 

\section{Variance reduction}

The method  described in the previous section has a scalar field $\phi$ 
for which each component (fixing space-time, colour and Dirac index) has
a typical variance of order 1. Thus the propagator will have  a standard
deviation of order $N_s^{-1/2}$ for $N_s$ samples of the $\phi$ fields.
The most promising way to improve on this situation is  to improve on
the   operator $\phi^*\phi$ used to calculate the stochastic estimators
of $G_{ij}$ in eq.~(\ref{stokinv}). Here we study in detail two
different methods and discuss their advantages and suitability for
effective implementation.

\subsection{Local multihit}

The easiest way to construct an operator that has a substantial reduction
in variance is to observe that is it possible to perform a local
multihit for the scalar fields $\phi$ needed for $G_{ij}$. This is
analogous to the method proposed in \cite{prr} for pure gauge systems
and is equivalent  to performing an average over infinitely many samples
of the chosen component of $\phi$ with all other components held fixed.
This has been proposed in ref.\cite{dfmp} and clearly leads to a
variance reduction.  Because of the simple quadratic nature of the
integration over $\phi$, the multihit average is obtained explicitly by 
 \be
  \phi_i \to -A_{ii}^{-1} A_{ij} \phi_j
  \ee
 with $i \ne j$ and no summation on $i$ and where, for the Wilson-Dirac
case,  the diagonal term is given by $A_{ii}=1+16 \kappa^2$. Thus each
$\phi$ field can be replaced by its multihit  average. It is permissible 
to use these multi-hit values in place of $\phi$  in evaluating
propagators and observables involving products of  $\phi$  fields
provided that  no  $\phi$ field is in the neighbourhood of another --- 
that is  no pair of $\phi$ fields can be linked by $A$ and so are
not nearest  or L-shaped next-to-nearest neighbours. 
 This multihit improvement is  easily implemented with only a minimal
effect on the computer time consumption and it provides a marked
improvement over no variance reduction. This improvement is independent
of the extent $T$ of the fermion propagation, however. Thus though the
improvement is substantial,  it does not allow a study of large $T$.
This is because the method   only averages over the  nearest and
next-to-nearest neighbours of each site, thus taking into account only
local variations in the scalar fields.

\subsection{Maximal variance reduction}

Instead of averaging over only near neighbours of a given site, one
could use all fields inside some given region $R$. Let $s_i$ be  the
scalar field variables at the boundary of $R$ and consider submatrices
of the matrix $A$: firstly  $\tilde{A}$ containing elements that link
the $\phi$ fields inside the region $R$ to those on the boundary and
secondly   $\bar{A}$ containing only links between the fields totally
inside the region $R$.  Now to average simultaneously over  all  scalar
fields inside $R$ while keeping the fields $s_i$ on the boundary fixed,
it is sufficient to replace $\phi$ at a given site $i$ with the average
obtained from the following expression:
 \be 
v_i = {1 \over \cal{Z}} \int D\phi \, \phi_i \exp \,
-{1\over 2} \left( \phi^*_j \bar{A}_{jk}\phi_k + \phi_j^*\tilde{A}_{jk}s_k +
s_j^*\tilde{A}_{jk}\phi_k   \right)  . 
 \label{maxvarred} \ee 
 Because the
integral (\ref{maxvarred}) is gaussian, one can easily calculate it
analytically to obtain 
 \be v_i = - \bar{A}^{-1}_{ij} \tilde{A}_{jk} s_k. 
 \ee
 where $i,\ j \in R$ and $k \not \in R$. We will call $v$ the variance
reduced estimator for $\phi$. By combining two  such improved
estimators, each from disjoint regions $R$ and $R'$ respectively, one
obtains a variance reduced estimator for propagator $G$ from any point
in $R$ to any point in $R'$. The choice of the two regions $R$ and $R'$
is arbitrary (subject to the constraint that the two regions should not
overlap in the sense of being linked by $A$). The local multihit
described above corresponds to taking  each region as just one site.
However, we can now optimise the choice of regions to obtain maximal
variance reduction.

In order to calculate $v$ in given gauge configuration, one needs the
inverse of $\bar{A}$ from an extended source -- the scalar field $s_i$
at the boundary of $R$. This is computationally equivalent to a single 
inversion of the Dirac-Wilson fermion matrix in region $R$. If the
volume of $R$ is large, this  is computationally demanding and so the
method is not immediately advantageous. The gain comes from the fact
that once this inversion is  done, one can  efficiently evaluate the 
propagators from every site inside region $R$ to every site inside
region $R'$. If the cost of calculating the necessary scalar
configurations is not too high, one should gain a substantial amount of
CPU time  compared to conventional methods. In addition the reduction in
variance should be much greater than for the local multihit method, as
one  averages the $\phi$ fields over a larger region. In the case of
fairly heavy  quarks, one can estimate analytically the variance
reduction using the hopping parameter  expansion: this gives a reduction
of $\kappa^d$ where $d$ is the minimum  number of links from the
boundary of $R$ to the interior point under consideration, and  likewise
for $R'$. Thus it is feasible that a stochastic evaluation of a hadronic
correlator involving  a separation of $T$ time steps will have its
variance reduced by $\kappa^T$. This achieves our goal of evaluating
efficiently large time propagators.

In a sense, both local multihit and maximal variance reduction are
three level Monte Carlo updating algorithms:
 \begin{enumerate}
 \item One generates gauge configurations $g$ with a suitable algorithm.
 \item In each $g$ one generates stochastic samples $\phi$ according to
distribution in eq.~(\ref{stokinv}).
 \item For each scalar field configuration one generates improved
operators keeping some of the original fields $\phi$ fixed. This can
be done analytically (or with Monte Carlo) for both maximal variance
reduction and local multihit.
 \end{enumerate}

 Since the last step can be performed analytically by an iterative
scheme,  the computational effort involves one inversion per stochastic
sample. Thus  for $N_s$ stochastic samples per gauge field, one will
have  to perform  roughly $N_s/12$ inversions compared to conventional
extraction of the  propagator from all colour-spins at one source point.
However, one gets access  to the propagator from all sources to all
sinks which may more than compensate. We now explore the implementation.

\section{Implementation}

To compare different methods, we have implemented the stochastic
inversion method, both with local multihit and with maximal variance
reduction.

If one is using unimproved Wilson fermions, writing the Monte Carlo
algorithm for scalar fields is straightforward. The only complication
arises from the fact that the action contains next-to-nearest-neighbour
interactions. To be able to  vectorize our algorithm in the style of the
conventional  red-black partition of odd and even sites, we assigned the
lattice sites to 32 ``colours'' and updated each colour sequentially. For
parallel machines such a partitioning is unnecessary. The actual heat
bath and overrelaxation algorithms are   simple. The local action,
obtained directly from eq.~(\ref{stokinv}) by keeping only terms involving
$\phi_{x}$ with others fixed, is just 
 \be
 S_{loc} =  {1 \over 2} \left( \phi_x^\dagger C \phi_x +\phi_x^\dagger
a_x+a_x^\dagger  \phi_x \right),
 \label{squad}
 \ee
with $C=1+16\kappa^2$ and 
 \be
a_{x} = -\kappa M_{xi}^\dagger\phi_i -\kappa M_{xi}\phi_i +
\kappa^2 M_{xi}^\dagger M_{ij} \phi_j,
\label{slocal}
 \ee
 where one should note that all sums over sites exclude the
$x$-site. Completing the square, then
 \be
S_{loc} =  {1 \over 2} \left( \phi_{x} + {a_{x}\over C}\right)^\dagger C
\left(\phi_{x} + {a_{x}\over C}\right),
 \ee
 and the heat bath algorithm is  equivalent to generating gaussian
random numbers with variance $C^{-1}$ and equating them to  $\phi_{x} +
a_{x}/C$. For the gaussian random numbers we use the Cray library
function SLARNV. The  overrelaxation is equally straightforward: one
just flips 
 \be
\phi_{x} \to \phi_{x} - {2 a_{x}\over C}
 \ee 
 for each Dirac and colour component of $\phi$.

In evaluating $a_x$, it is very inefficient to  use the matrix $A$
directly since it connects  54 sites to $x$. As in eq.~(\ref{slocal}),
using the result that $A=(1-\kappa M)^{\dagger} (1-\kappa M)$, it is
preferable to work with $M$ directly since it only  has an implicit sum
over 8 sites. Then the main computational load  in evaluating $a_x$
comes from the gauge part of the matrix multiplication  $M\phi$. If one
keeps $\psi=M \phi$ in  memory as well as $\phi$ itself, then the
evaluation of $a_x$  from $M^{\dagger}(-\kappa \phi +\kappa^2
\psi)-\kappa \psi$ involves  only one application of $M^{\dagger}$ to a
vector. One then needs, however, to  update $\psi$ which involves work
equivalent to a further application of $M$ to a vector. This strategy 
reduces the total work needed to the equivalent of two applications 
of $M$ to a vector.

In practice we found that,  after initialising using  heatbath sweeps,
it was efficient to use combined sweeps of  4 overrelaxation plus one
heatbath to give sufficiently equilibrated and independent samples. We 
discuss the number of such sweeps in detail later.   Where independence
of the samples is at a premium, one can choose to combine only samples
further apart - this we explored and we report later on the result. In
general, as one approaches the  chiral limit of light quarks, one
expects the fermion matrix $A$ to have small eigenvalues  with spatially
extended eigenvectors. These will cause critical slowing down of our
local  updating scheme. Similar considerations apply to using bosonic
algorithms for dynamical fermions~\cite{lf} and multi-grid and other
methods are  known to be available to circumvent this problem in
principle.

For the Sheikholeslami-Wohlert improved clover action \cite{sw} the
algorithm is not much more complicated. The Dirac-Wilson fermion matrix
$Q$ is replaced by
 \be
Q_{SW} = L - \kappa M \ ,
\ee
 where $L$ is  diagonal in space-time but not now in colour and spin and
depends on the coefficient $c_{SW}$ which is 1.0 in lowest order
perturbation theory  and, as discussed later,  can be estimated by
tadpole improvement or non-perturbative improvement:
 \be
L= 1 - {c_{SW}\over 16}\kappa \sum_{\mu\nu} F^{\mu\nu} \sigma_{\mu\nu}\ ,
\label{cloverL}
 \ee
  where $F_{\mu \nu}$ is defined here at each site as the lattice
operator  $\sum_P (U_P - U_P^\dagger)$ given by the 4-leaved clover sum 
over  plaquettes on the lattice in the $\mu,\ \nu$ plane where $U_P$ is
the product of the four links around a plaquette $P$ and where 
 \be
\sigma_{\mu\nu} = {1\over 2} \left[ \gamma_\mu, \gamma_\nu \right]\ .
 \ee
 Note that $L$ is hermitian. 

The local action is still quadratic in the  $\phi$ field at a  site
$x$ as given by eq.~9\ref{squad}), but $C$ is now a matrix in Dirac
and colour indices:
 \begin{eqnarray}
C &=& \left(1 - {c_{SW} \over 16}\kappa F^{\mu\nu}
\sigma_{\mu\nu}\right)^\dagger \left(1 - {c_{SW} \over 16}\kappa F^{\mu\nu}
\sigma_{\mu\nu}\right) + 16\kappa^2, \\
 a_x &=& -\kappa \left(1 - {c_{SW} \over 16}\kappa F^{\mu\nu}
\sigma_{\mu\nu}\right)^\dagger M_{xi} \phi_i
-\kappa M_{xi}^\dagger \left(1 - {c_{SW} \over 16}\kappa F^{\mu\nu}
\sigma_{\mu\nu}\right) \phi_i \nonumber\\
 & & + \kappa^2 M_{xi}^\dagger M_{ij} \phi_j \ .
 \end{eqnarray}
 For updating a given colour-spin component of  $\phi$, we only need the
inverse of the appropriate real diagonal element of $C$. However, the
non-diagonal  terms in $C$ need to be added to the force term $a_x$. 
With these changes,  the actual updating algorithms for the clover
action are the same as for unimproved Wilson fermions. As well as
storing intermediate results ($M\phi$ and $L\phi$)  to save computation
as described for the Wilson case, the clover term can be  treated
efficiently by noting~\cite{stephenson} that projecting $\phi=\phi_+ +
\phi_-$, where $\phi_\pm={1 \over 2}(1\pm \gamma_5) \phi$, allows $L$
and $C=L^\dagger L$  to be represented as two $6 \times 6$ hermitian
matrices rather than one $12 \times 12$ matrix at each site.

In addition to Monte Carlo algorithms for scalar fields one needs an
iterative inversion algorithm in region $R$ to implement the maximal
variance reduction with source $\tilde{A}_{jk}s_k$. Since the matrix
$\bar{A}$ is hermitian, a reliable method  is  conjugate gradient
and this is what we use. Since the condition number of
$A=Q^\dagger Q$ is considerably worse than that of $Q$ itself,  it may
well be faster to use a method such as minimal residual to invert the
non-hermitian  matrix $Q$  and then in turn  to invert  $Q^\dagger$,
particularly if an efficient preconditioner such as red-black can be
used for  these inversions. Since the present study is exploratory, we
have not investigated this option further. Another option is that since
we need to invert  $\bar{A}$ for the same gauge configuration with $N_s$
different sources $s_k$ coming from the stochastic  Monte Carlo described
above, inversion methods using multiple sources may  offer some
computational benefit.

We now discuss the optimum choice of the partitions $R$ and $R'$ for 
applications. Sites in $R$ and $R'$ must not be connected by $A$. The 
matrix $A$ contains nearest and next-to-nearest link terms. Because  of
the spin projection ${1 \over 2} (1 \pm \gamma_\mu)$ contained  in the
$\pm \mu$ -directed link term of the Wilson fermion matrix $M$,   $A$
does not contain any terms with double straight links. Thus a  simple
way to divide $R$ and $R'$ is by a time-plane on which the sources 
$s_k$ lie. Since the lattice is periodic in time (or antiperiodic for
fermions),  the optimum situation is to have two such boundaries, for
example  at $t_1=0$ and $t_2=T/2$ where $T$  is the time extent of the
lattice. Then any propagator from the region  $0 \le t \le T/2$ to the
region $ T/2 \le t \le T$ can be evaluated. Note that propagators from
one region to the source area $S$ are  allowed and will be variance
reduced. A propagator  entirely within one region will involve two
$\phi$  fields (say at $x$ and $y$) in that region and the integration
over the fields inside $R$  of  eq.~(\ref{maxvarred}) will then give an
extra  disconnected term involving $\bar{A}^{-1}_{xy}$. This is just a
propagator within region $R$ and so we are back to the problem of
evaluating it for  all pairs of points $x$ and $y$ in $R$. Thus our
present method does not allow  any  variance reduction for a propagator
corresponding to a disconnected fermion loop.

 In applications, we  create $N_s$ independent samples of the scalar
field $\phi(x)$ for each gauge configuration. We then use the  $\phi$
field as a source on time planes $t_1$ and $t_2$ to obtain the  
variance reduced fields $v(x)$ for each sample in $R$ and $R'$. As well
as $v$ in $R$ and $R'$, we then only need  to store $\phi(x)$ on the two
source time-planes (which we call region $S$). So each variance-reduced 
sample has storage of $24 \times L^3 \times T$ real numbers which is 
equivalent to one twelfth of the storage of the usual propagator from
one point  to all sinks.
 These variance reduced fields then allow improved estimators of the 
propagator from any point in $R+S$ to any point in $R'+S$. This allows 
a determination of hadronic correlators involving one light quark using 
nearly all points as sources and sinks. We will investigate whether the 
increase in statistics from using so many source points is sufficient to
 compensate for the stochastic noise inherent in the method.  

We now discuss the choice of the number of samples $N_s$. If too many
samples  were used, the determination of the correlator of interest
might have a  variance from one gauge configuration which is smaller
than the variance over many gauge configurations. In other words, there
will be no advantage in  measuring too accurately on one gauge
configuration. 
 For correlators involving one light quark, the partition of
computational effort  between more samples $N_s$ per gauge or more gauge
configurations is not  crucial. Provided one does not overdo $N_s$ as
describe above, the signal  should be comparable for a given product of
$N_s$ and number of gauge configurations.

 When more than one light quark propagator is to be evaluated
stochastically in an unbiased way,  the considerations of optimum $N_s$
are more subtle. Provided the scalar field samples  are independent, the
two light quark propagators, each from $R+S$ to $R'+S$, can be estimated
from $N_s^2$ combinations of the samples on each gauge configuration
(${1 \over 2} N_s (N_s-1)$ combinations if both light  quarks have the
same mass so are taken from the same set of samples). This suggests that
 the noise on the combined signal may decrease as fast as $N_s^{-1}$ in
this  case. This would imply that larger values for $N_s$ were more
efficient  in this case. We will report on our investigation of this
point.

 For studies of baryons or of matrix elements involving mesons, three or
more  light quark propagators are needed. Provided no propagator lies
entirely  within one of the regions $R$ or $R'$ this is feasible. For
mesonic matrix elements,  one way to achieve this is to put the matrix
element insertion on the source time-plane  ($S$).

\section{Static systems}

We have chosen the system containing one light and one infinitely heavy
quark (static quark) as our main test case. This system describes the
$B$ meson in leading order heavy quark effective theory. With
conventional  light quark inversion techniques, the propagator from one
source only is evaluated and this allows the hadronic correlator to be
obtained from only  two sink locations on a given gauge configuration for a
given $T$.  This makes very little use of the information contained in
the gauge field. In contrast, the stochastic approach allows the
hadronic propagators to be determined  from very many more sites.
Furthermore since the hadronic observable is linear  in the light quark
propagator, any problems of correlations among the statistical  samples
of the $\phi$ fields are irrelevant. This is thus an optimum area  for
testing the stochastic method. Indeed previous work~\cite{dfmp} 
using multi-hit improvement has already concentrated in this area. Here 
we  compare our maximal variance reduction approach with this approach 
and also with the  conventional iterative inversion. Our main point of
comparison will be   the B-meson correlation at $T=7$. 

Because of the flexibility of the stochastic method, it is possible to 
study non-local hadronic operators with no additional computational
effort. Since orbital excitations involve non-local operators, this
allows a comprehensive study of the excited state spectrum of
heavy-light mesons. This is an area where comparatively little is known,
so we are able to show the power of our approach by determining several
new features of the excited B meson spectrum.   We also explore the
baryonic  spectrum in the static limit and report on the comparison with
other lattice work and with experiment.

\subsection{B meson in the static limit}

Following the conventions of \cite{lm}, we use nonlocal operators for
the B meson and its excited states. This will enable us to  study also
the orbitally excited mesons - the details are collected in the Appendix.
The operator $B$ we use to create such a $\bar{Q} q$ meson on the
lattice is defined on a timeslice $t$ as  
 \be B_t = \sum_{x_1, x_2} \bar{Q}_({\bf x}_2,t) 
    P_t({\bf x}_1, {\bf x}_2) 
    \Gamma 
   \, q({\bf x}_1,t)\ .
\label{bmeson}
 \ee
 $Q$ and $q$ are the heavy and light quark fields respectively, the sums
are over all space at a given time $t$, $P_t$ is a linear combination of
products of gauge links $U$ at time $t$ along paths $P$ from ${\bf x}_1$
to ${\bf x}_2$, $\Gamma$  defines the spin structure of the operator.
The Dirac  spin indices and the  colour indices are implicit. The masses
are then calculated from the exponential fall off of the $B\bar{B}$
correlation function  (or vacuum expectation value)
 \ba C(T) &=& \left< B_t \bar{B}_{t+T}\right>_0 \\
     &=& \left< \bar{Q}({\bf x}_2,t) 
                P_t({\bf x}_1, {\bf x}_2) 
                \Gamma
                q({\bf x}_1,t) \right. \times \nonumber  \\
     & &
                    \left. 
                    \bar{q}({\bf x}'_1,t+T)
                    P_{t+T}({\bf x}'_1, {\bf x}'_2) 
                    \Gamma^\dagger 
                    Q({\bf x}'_2,t+T) \right>_0 \\
     &=& \mbox{Tr } \left< 
                  P_t \Gamma G_q({\bf x}'_1, t+T, {\bf x}_1, t) 
                  P_{t+T} \Gamma^\dagger G_Q({\bf x}_2, t, {\bf x}'_2,t+T)
                     \right>_0 \ .
    \label{corrB}
 \ea
 We have denoted the light and heavy quark propagators by $G_q$ and
$G_Q$ respectively and the trace is over Dirac and colour indices and
also includes the spatial sums over ${\bf x}_1$, ${\bf x}'_1$, ${\bf
x}_2$ and ${\bf x}'_2$. Because we work with  static  heavy quarks, 
up to an irrelevant overall constant,  one has
 \be
G_Q({\bf x}_2, t, {\bf x}'_2 ,t+T) = {1\over 2}(1+\gamma_4) 
 U^Q({\bf x}_2,t,T) \delta_{{\bf x}_2, {\bf x}'_2}\ ,
\label{hqpropag}
 \ee
 where the gauge link product for the heavy quark is
 \be
 U^Q({\bf x},t,T)=\prod_{i=0}^{T-1} U_4({\bf x},t+i)\ .
 \ee
 Now for the light quarks, we wish to evaluate the propagator $G_q$ by
stochastic  methods using eq.~(\ref{propagator}) where now angle brackets
refer to  the average over the $N_s$ stochastic samples.  An alternative
form can be obtained for the Wilson-Dirac discretisation, for which
$Q(x,y)=\gamma_5 Q(y,x)^{*} \gamma_5$ where $x$ includes space-time,
colour  and spin labels. This is 
 \be
G_{ji}=  \gamma_5 \langle (Q_{jk} \phi_k) \phi_i^* \rangle \gamma_5 \ .
\label{bmeson-alt}
 \ee
 In practice, we find it optimum to evaluate both of these expressions 
eq.~(\ref{propagator}) and eq.~(\ref{bmeson-alt}) using our stochastic
estimators and average them.  Since we shall need it frequently,  we
define $\psi_i = Q_{ij} \phi_j$.

Using eq.~(\ref{hqpropag}) for the heavy quark propagator and stochastic
scalar fields according to eq.~(\ref{propagator}) for the light quark
propagator in eq.~(\ref{corrB}) one gets
 \be
 C(T) = \mbox{Tr } \left< P_t \Gamma \phi({\bf x}'_1, t+T) \psi^*({\bf
x}_1, t))      P_{t+T} \Gamma^\dagger {1 \over 2} (1+\gamma_4)
 U^Q({\bf x}_2,t,T)
                  \right> \ .
\label{corr-bmeson}
 \ee

By choosing different path combinations and appropriate choices of
$\Gamma$ in eq.~(\ref{bmeson}) one can obtain different $J^P$ states as
described  in the Appendix.
  For the ground state B mesons in the static limit, we will have a 
degenerate pseudoscalar and vector. (The splitting between them can be 
evaluated by taking matrix elements of the clover term $\sigma_{\mu \nu}
F_{\mu \nu}$ in the B ground state). The simplest hadronic operator to
create these states is then  obtained from eq.~(\ref{bmeson}) by
choosing $P_t=1$ and $\Gamma=\gamma_5$ for the pseudoscalar and
$\Gamma=\gamma_i$  for the vector. From eq.~(\ref{corr-bmeson}) one
obtains
 \be
    C(t)= {\rm Tr }\  H_{-}\ \left< \phi(x) 
 U^Q({\bf x},x_4,t)   \psi^*(x+\hat{4}t)  \right>\ ,
 \ee
  where $H_\pm=(1\pm\gamma_4)/2$. The sum over Dirac indices is very
simple with our convention in which $\gamma_4$ is diagonal and, making
the colour sums explicit too, this is  
 \be
    C(t)= \sum_x \sum_{i=3,4} \left< \phi_{ai}(x) 
 U^Q_{ab}({\bf x},x_4,t) \psi_{bi}^*(x+\hat{4}t)  \right> \ .
 \ee The alternate expression (\ref{bmeson-alt}) for the stochastic
light quark propagator yields
 \be
    C(t)=  \sum_x \sum_{i=1,2} \left< \psi_{ai}(x) 
 U^Q_{ab}({\bf x},x_4,t) \phi_{bi}^*(x+\hat{4}t)  \right> \ .
 \ee

If we now want to use maximally variance reduced operators instead of
$\phi$ and $\psi$, one has to pay attention to the fact that $\phi$ and
$\psi$ must not come from the same  partition. Here we choose the
simplest partition with sources at time planes $t_1$ and $t_2$. One
subtlety is that $\psi=Q\phi$ is smeared out by one link in each
direction:  thus $\psi(t_1) \phi(t_1+t)^*$ with $t > 0$  would be
invalid.  Our favoured setup (which we tested to have minimal variance
and which corresponds  naively to taking the estimators as far as
possible from the source) is to take $\psi$  at $t_1 \pm (t/2)_+$ and
$\phi$ at $t_1  \mp (t/2)_-$ where if $t$ is  odd $(t/2)_+$ is rounded
up, etc. So for $t=1$ we have $\psi(t_1\pm1) \phi(t_1)^*$. Since the
drop-off of $\phi$ from the source at $t_1$ is  roughly exponential up
to half way between the source time-slices, other partitions of $t$
have variance which is not  much greater than our favoured setup. The
complication is how to combine efficiently several estimators which
have somewhat different variance.

To improve the overlap of our operators with the ground state, we have
also considered fuzzed operators. These have paths $P_t$ formed by
joining the light and heavy quarks by straight links of  length $l$ in
all 6 spatial directions. These  links are themselves spatially
fuzzed~\cite{fuzz} using an iterative scheme.  We use two different 
lengths $l$ each with a different number of fuzzing iterations as well
as the  unfuzzed operators described previously. Correlations of all
combinations at each end are evaluated giving us a $2 \times 2$ or $3
\times 3$ matrix.

\subsection{Comparison of methods}

 Since there is a considerable body of data on B meson correlations for 
Wilson fermions, we first tested our approach with Wilson fermions.

 For comparison purposes, we considered a small lattice ($8^3$ 16) at 
$\beta=5.74$ with Wilson fermions of hopping parameter $K=0.156$. This
choice was motivated by pre-existing studies~\cite{wuppertal}. As a first
example we  evaluated the B meson correlator at time separation $t$
using local hadronic  operators at source and sink. Then we compared
conventional inversion with  various implementations of  stochastic
inversion. Results for the correlator $C(7)$ at $t=7$ are shown in
Table~\ref{table:comparison} where the comparison has been made  for
equal disk storage of propagators or scalar field samples.

For the stochastic inversion methods we used 20 gauge configurations,
each containing 25 samples of the scalar field. For the gauge fields we
use 100 combined sweeps of one Cabibbo-Marinari  pseudo-heatbath
algorithm  followed by 3 overrelaxation  steps between configurations. 
The scalar fields were evaluated as described previously  by using 125
heatbath plus overrelaxation updates between measurements after 250
sweeps to thermalise the first sample for each gauge configuration. We
tested that our results were unchanged if more thermalisation  sweeps
were used.  For the conventional MR inversion method we used 4 different
sources on 10 gauge configurations. Since the storage  of the
conventional propagator from one source involves 12 colour-spins, it is
the same as the storage of 12  stochastic scalar fields, so the
comparison is made at equal file storage.

\begin{table*}[hbt]
\setlength{\tabcolsep}{1.5pc}
\newlength{\digitwidth} \settowidth{\digitwidth}{\rm 0}
\catcode`?=\active \def?{\kern\digitwidth}
\caption{B meson correlators at $t=7$.}

\begin{tabular*}{\textwidth}{@{}l@{\extracolsep{\fill}}llcc}
\hline
Method & $C(7) \times  10^7$ & Data Set &  CPU \\
\hline
MR inversion  & 3712(147) & propagators from 4 sources  & 1\\
 & & for 10 gauge fields & \\
\hline
Stochastic inversion & & 25 samples of $\phi$ &  \\
\ Basic & 2754(926) & for 20 gauge fields & 2 \\
\ Local multihit & 3418(410) &    &2 \\
\ Maximal variance reduction & 3761(21) &  & 4 \\
\hline

\end{tabular*}
\label{table:comparison}
\end{table*}

\begin{figure}[t!]
\vspace{11cm}
\includegraphics{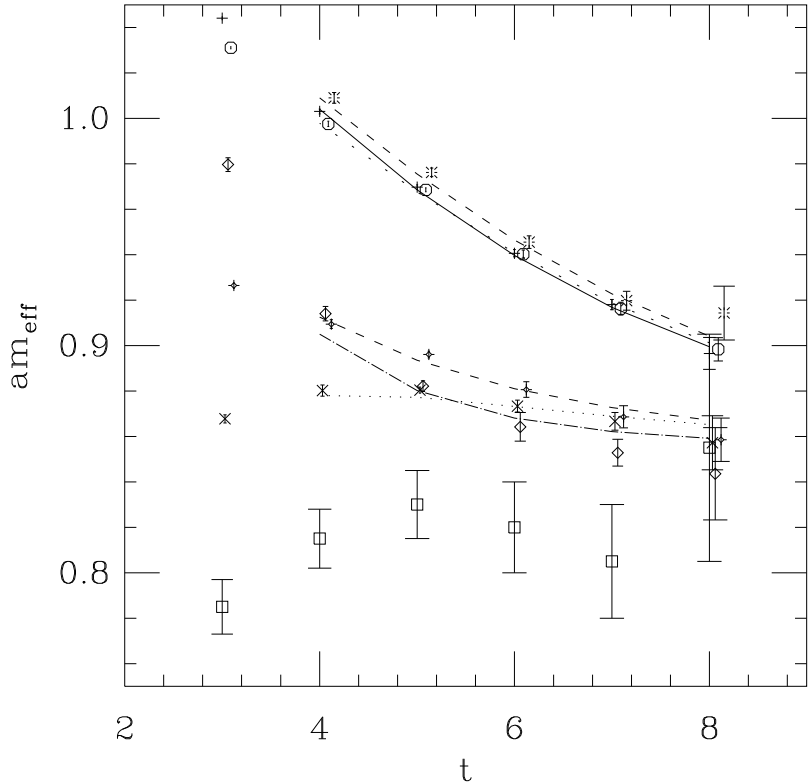}

\caption{  The  B meson effective mass  versus $t$ from our data at
$\beta=5.74$ from $8^3 \times 16$ lattices with  Wilson fermions and
different combinations of  local, fuzzed and $\gamma_i D_i$ sources and
sinks  together with a three exponential fit.
 Also shown (squares) are the Wuppertal data~{\protect\cite{wuppertal}}
for smeared source and local sink from 170 propagators.}

\end{figure}

Clearly the maximal variance reduction gives a factor of 7 improvement
in error  for only an overall computational increase of a factor of 4.
This is equivalent to  a net gain of a factor of 12 in computing time
for a similar result. Moreover, the stochastic method allows 
correlations involving different sources (smeared, fuzzed, orbitally
excited, etc) to be constructed at little  extra cost. This is shown in
fig.~1 where a comparison is made  of our results for the effective mass
with results~\cite{wuppertal} from conventional inversions  (with 170
propagators) which are seen to be significantly less precise than those
obtained here in the $t$ region of interest. More details of the fit are
presented in ref.~\cite{tsukuba}.

 In order to explore more fully the power of the method of maximal
variance reduction of stochastic propagators, we then undertook a more
extensive  study using clover improved fermions. In this case the focus 
of attention was on the precise determination of the B meson spectrum
and excited states.

\subsection{The excited B meson spectrum}

Our results are not only useful for comparing different methods but are
also physically interesting in their own right. In particular, the
spectrum of the exited states in the static limit  has not been thoroughly
studied and experimental information on this spectrum is also limited. 
Moreover, without a precise extraction of the excited state component in
the S-wave,  the ground state contribution will be uncertain which implies
systematic errors in  extracting heavy-light matrix elements (for
example $f_B$).

In order to study these new areas of physics using the power of 
stochastic inversion with maximal variance reduction, we determine the
spectrum of heavy-light mesons and their excited states in the static
limit. In order to minimise order  $a$ effects, while still keeping in
contact with existing simulations, we have  used a tadpole improved
action at $\beta=5.7$. A non-perturbatively improved~\cite{NP} action 
is preferable to the tadpole-improved prescription on theoretical
grounds but the determination  of the appropriate value of the clover
coefficient $c_{SW}$ has not been feasible for $\beta \le 6.0$. The
results at larger $\beta$ than 5.7  do, however, suggest that the
non-perturbative value for $c_{SW}$ would be significantly  larger  than
the tadpole value we use here.  We also wish to keep  finite size
effects under control so we use two spatial lattice sizes.

We have performed simulations on $8^3\times 16$ and $12^3\times 24$
lattices with $\beta=5.7$ with $c_{SW} = 1.57$ and we study two
different values of hopping parameter: $\kappa_1 = 0.14077$ and
$\kappa_2=0.13843$. These values have been used before to study the
effect of tadpole improvement on the  light meson spectrum~\cite{scale}
and  pseudoscalar meson and vector meson masses are available from that
work (see also Table~\ref{table:PV}).  The chosen light quark masses
correspond roughly to the strange quark mass ($\kappa_1$) and to twice
the strange quark mass ($\kappa_2$). We will describe our light quark
masses in dimensional units by  quoting $(r_0 m_P)^2$  where
$r_0/a=2.94$ is used at $\beta=5.7$ from our own interpolation.  A
recent independent study~\cite{r0scri} gave $r_0/a=2.99(3)$ at this
$\beta$ value. We also use $r_0$ to set the scale of the B meson 
masses. In terms of conventional units, $r_0 \approx 0.5$\ fm.

For the 20 pure gauge configurations we use a conventional scheme with 
200 combined sweeps of 3 overrelaxation plus one heatbath between 
configurations.   We evaluate $N_s=24$ scalar field samples per gauge
configuration using 25 combined sweeps of 4 overrelaxation plus one
heatbath, after 125 heatbath sweeps to initialise from a cold start. In
each case we then evaluate the  variationally improved scalar fields
using conjugate gradient in the regions between  time slices 0 and
$T/2$. For the hadronic operators we use spatial fuzzed links which are
iteratively evaluated~\cite{fuzz} by summing ($f \times$ straight + sum
of 4 spatial U-bends) and projecting  the result to SU(3). Using
$f=2.5$, we choose two fuzzed superlinks: (i)  2 iterations of fuzzing
with superlinks of length 1, and (ii) 8 iterations  of fuzzing with
superlinks of length 2. When we explore  Bethe Saltpeter wavefunctions
for B mesons, we also employ other lengths for superlinks.

\begin{figure}[t!]
\vspace{12cm} 
\includegraphics{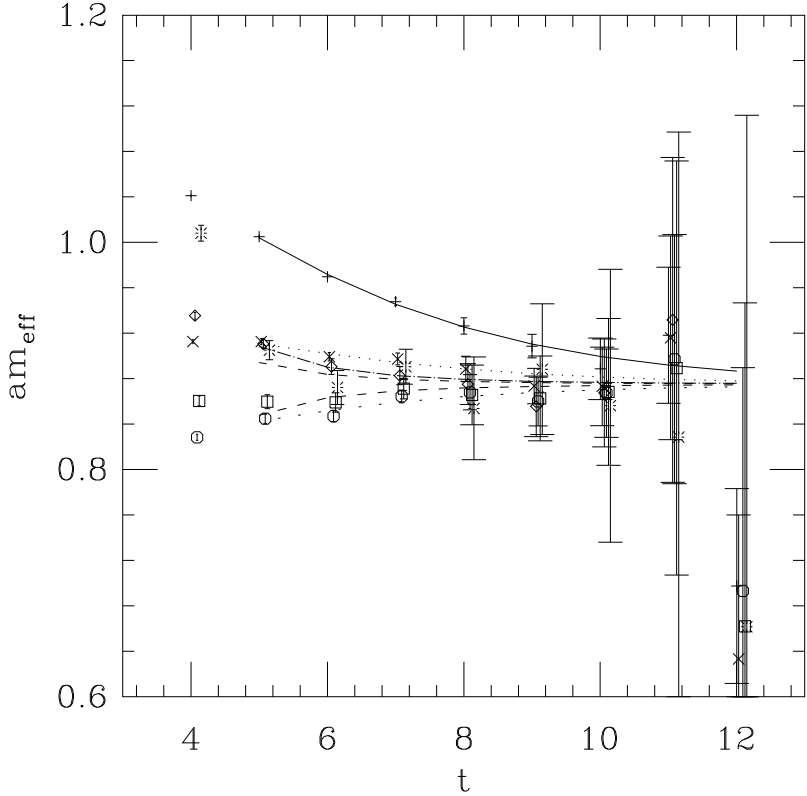}
  \caption{\protect Effective mass plot for the $S$-wave $\bar{Q}s$
meson using clover fermions on $12^3 \times 24$ lattice at $\beta=5.7$
with light quark hopping parameter $\kappa_1=0.14077$. } 
  \label{figure:effmass}
\end{figure}

 Our basic method for extracting the mass spectrum is to fit the  matrix
of zero-momentum correlators at a range of time separations to a 
factorising sum of several states. We use either two states or three,
and  in the latter case we may fix the mass of the third state to 2.0 in
lattice units to stabilise the fit.  A typical effective mass plot can
be seen in Figure~\ref{figure:effmass}, where we have plotted the
effective mass of the $L=0$ (S) state together with a factorising fit.
We use either uncorrelated fits or some model of the
correlation~\cite{amck}.  Typically the modelled correlation (this is
the correlation among measurements from different gauge samples) is used
to find  the $t$ range giving acceptable fits. Then an uncorrelated fit
is used  to give the central values of the masses and other parameters. 
Statistical errors are determined by bootstrap of the gauge
configurations. The systematic errors from  fitting are estimated by
varying the fit range in $t$ and the fit correlation model - these
systematic errors are only quoted if they  are significantly larger than
the statistical errors.

\begin{table}[hbt]

\caption{B meson effective masses.}

\begin{center}
\begin{tabular}{lllllll}


  state &$\kappa$&L& $am$     & $am'$      & $\chi^2$/dof & $t$ range \\
\hline
S      &1&12 & 0.875(06) & 1.271(10) & 32/54-11 & 4-12 \\
P$_-$   &1&12 & 1.214(43) & 1.727(53) & 28/60-11 & 3-12 \\
P$_-$    &1&12& 1.194(48) & 1.697(54) & 13/30-8  & 3-12 \\
P$_+$    &1&12& 1.262(56) & 1.698(57) & 20/30-8  & 3-12 \\
D$_\pm$  &1&12& 1.555(12) & 1.825(35) & 22/27-6  & 4-12 \\
D$_-$    &1&12& 1.423(20)& 1.751(27)& 41/30-6  & 3-12 \\
D$_+$    &1&12& 1.744(56)& 2.039(61)& 15/30-6  & 3-12 \\
F$_\pm$  &1&12& 1.850(36) & 2.053(44) & 28/30-6  & 3-12 \\
\hline
S      &1&8& 0.877(26) & 1.273(44) & 17/30-11 & 4-8 \\
P$_-$  &1&8& 1.200(90) & 1.647(73) & 12/18-8  & 3-8 \\
P$_+$  &1&8& 1.222(120)& 1.774(70) & 20/15-8  & 3-7 \\
\hline
S       &2&12& 0.912(06) & 1.284(10) & 35/54-11 & 4-12 \\
P$_-$   &2&12 & 1.313(17) & 1.797(32) & 77/60-11 & 3-12 \\
P$_-$    &2&12& 1.329(19) & 1.809(41) & 41/30-8  & 3-12 \\
P$_+$    &2&12& 1.386(27) & 1.823(24) & 21/30-8  & 3-12 \\
D$_\pm$  &2&12& 1.578(10) & 1.826(30) & 38/21-6  & 3-10 \\
D$_-$    &2&12& 1.480(13) & 1.773(18) & 34/30-6  & 3-12 \\
D$_+$    &2&12& 1.710(43) & 1.883(45) & 28/30-6  & 3-12 \\
F$_\pm$  &2&12& 1.901(24) & 2.102(54) & 53/30-6  & 3-12 \\
\hline
S        &2&8& 0.899(12) & 1.290(21) & 11/30-11 & 4-8 \\
P$_-$   &2&8& 1.263(50) & 1.837(48) & 12/18-8  & 3-8 \\
P$_+$   &2&8& 1.224(71) & 1.721(49) & 5/18-8   & 3-8 \\
\hline

\end{tabular}
\label{table:meff}
\end{center}
\end{table}

 Our results for the masses are collected in Table~\ref{table:meff}.

Here the  different operators used correspond to those defined in the
Appendix. The two  values quoted for the $P_-$ state correspond to (i)
using the same fit as for the $P_+$ state  to yield the mass difference
most reliably and (ii) using the extra operators available for the $P_-$
case to get the best mass determination.  We determined the mass
difference of the $P_-$ and $P_+$ using a bootstrap analysis of this
difference and  obtain, from the $12^3$ spatial lattices, values of the
mass difference in lattice units of 0.068(64) at $\kappa_1$. The $P_+$
state for strange light quarks is thus heavier than $P_-$ with a
significance of 1 standard deviation. For the D-wave states, we  find
for the mixed operator (labelled $D_{\pm}$) a  mass consistent  with
lying between the masses of the two states separately. The splitting
between the  $D_-$ and $D_+$ masses appears to be quite large

\begin{figure}[t!]  
\vspace{10cm} 
\includegraphics{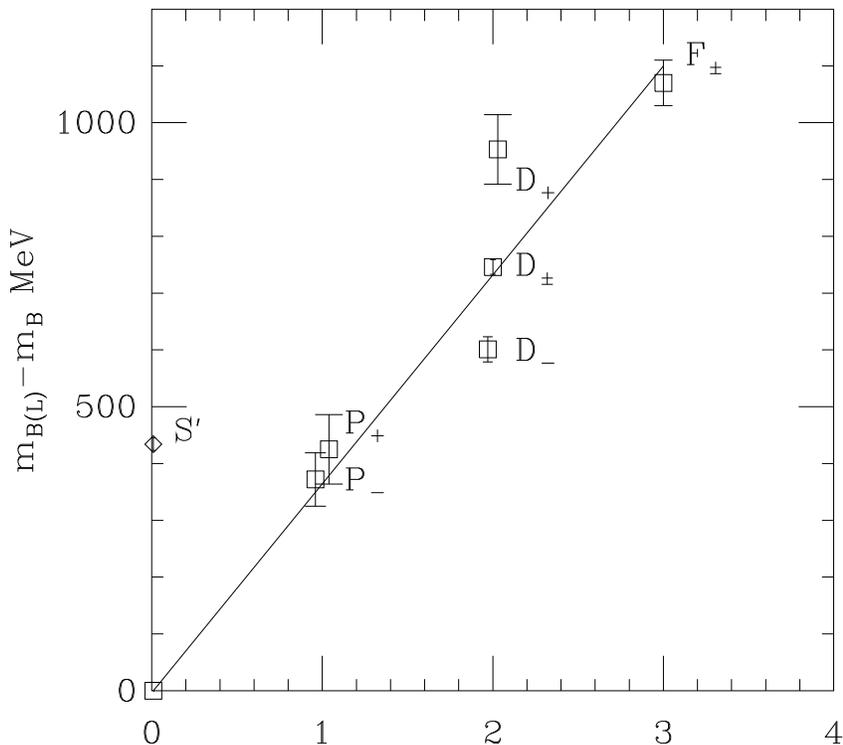}
 \caption{ The masses of excited $Q\bar{s}$ mesons  versus angular
momentum $L$ from  clover fermions with $\kappa=0.14077$. For the $L=2,\
3$ states, results from operators which are a mixture of  the two levels
are also plotted. The straight line is to guide the eye. The scale is
set by $a(5.7)=0.91 $GeV$^{-1}$.
   }
 \label{figure:SPDF}
\end{figure}

The absolute  values of the masses obtained in the static limit are not
physical because  of the self-energy of the static quark. We present
masses by taking the difference  with the ground state S-wave state (the
usual B meson). The dependence on the orbital angular momentum is shown
in Fig.~\ref{figure:SPDF} for strange quarks ($\kappa_1$). This suggests 
that the energy of the orbital excitations is linear with angular momentum.

\begin{figure}[t!]
  \centering
  \leavevmode
  \epsfysize=10cm
  \epsfbox{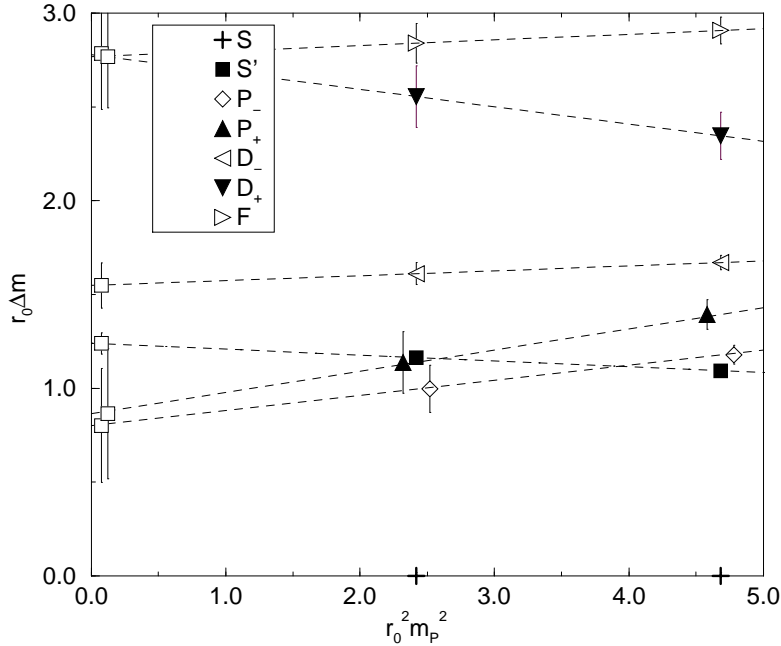}
  \caption{\protect The spectrum of B mesons containing one light (with
mass proportional to pseudoscalar mass $m_P^2$) and one infinitely heavy
quark. Masses are given in terms of $r_0$. The straight lines show a linear 
chiral extrapolation.
 } 
  \label{figure:mB}
\end{figure}

The dependence on  the light quark mass (through $\kappa$) would be
expected to be small  since the effect should be similar for each state
and so cancel in the difference. Our results from the larger lattice 
are broadly compatible with this picture - see Fig.~\ref{figure:mB} where 
our results from $\kappa_1$ and $\kappa_2$ are plotted.

 We see evidence of significant finite size effects in comparing our 
results at $L=8$ and $L=12$. Because of this,  we do not show  results
in Table~\ref{table:meff} from our smaller spatial lattice for the
higher lying excitations where the effect of the finite spatial size 
could be even larger.
 One specific example of the finite size effects  is that, for
$\kappa_2$, the $P_+$  state appears lighter than $P_-$ for $L=8$ while
the order is the other  way around at $L=12$, although the statistical
significance  of these level orderings is limited.  This order of the
$P_\pm$  levels  at $L=8$ was also found in  our results~\cite{tsukuba} 
from Wilson fermions. This is an issue of direct physical interest since
potential models indicate that the long range spin-orbit  interaction
can, in principle, yield a $P_+$ state lying lighter than the $P_-$ as
the light quark mass decreases to zero. Our larger volume results do not
support  this scenario and we find the $P_-$ state to lie lower than
$P_+$ with a  significance of 1 $\sigma$. To explore this in more
detail, we need to establish  the effect of the finite size effects.
This is especially relevant for excited states, since they would be
expected to be more extended spatially. This can be studied by
determining the wavefunctions of the various states.

\begin{figure}
  \leavevmode
  \centering
  \epsfysize=8.5cm
  \epsfbox{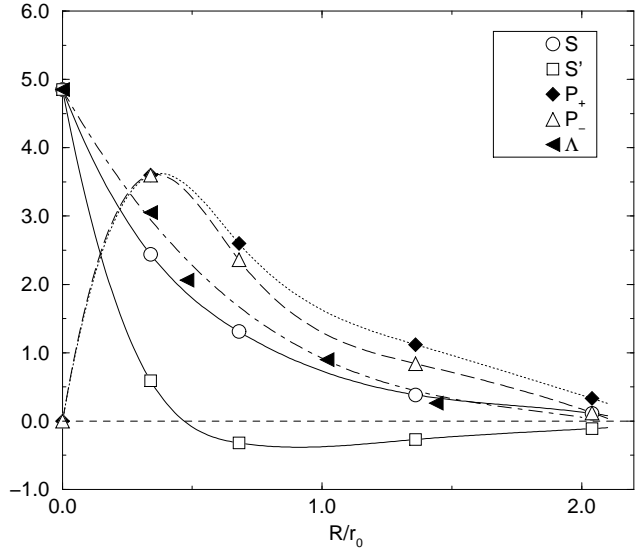}
  \vspace{-1cm}
  \caption{\protect The Bethe Saltpeter wave function of the $\bar{Q}q$
mesonic states. The radius $R$ of the light quark $q$ from the static
quark $Q$ is given in units of $r_0$. Here the light quark mass
corresponds to $\kappa_2$. For  the $\Lambda_b$ baryon,  the results
shown are obtained as described in the text. The continuous lines are to
guide the eye. 
 }
  \label{figure:wf}
\end{figure}

 We can determine the Bethe Saltpeter wave functions $w(R)$ of the B
meson states directly by fitting the ground state contribution (of the
form  $w(R_1)w(R_2)\exp(-mt)$) to  a hadronic correlator where the
operators at sink and source are of spatial size $R_1$ and $R_2$
respectively. Thus we measure  correlations for a range of  spatial
extents $R$ of the lattice operators used to create and destroy the
meson. We explore this at our larger lattice volume. In practice,
following~\cite{fuzz}, we use  straight fuzzed superlinks of  length $R$
(we keep the number of iterations of fuzzing fixed at 4 here). After
fitting, we extract the wavefunctions which are plotted for light quark
mass corresponding  to $\kappa_2$  in Fig.~\ref{figure:wf}. This clearly
shows the expected behaviour of higher orbital excitations  being more
spatially extended. Moreover, it shows evidence that the $P_+$ state is
fatter than the $P_-$ which could explain the mass difference dependence
on volume noted above - namely that  the $P_+$ state has a  modified
mass at $L=8$. Changing the  light quark mass from $\kappa_2$ to
$\kappa_1$ results in no change  in the wavefunctions within errors.

 Another comforting conclusion is that there is evidence that the
excited  S-wave state has a node at radius $R \approx 1.5 a$ which 
corresponds to $0.5 r_0$. This implies  that the meson has a nodal
surface with a diameter of approximately $r_0$ which is broadly
compatible with the result~\cite{fuzz} for S-wave mesons made from two
light quarks that the node for the excited state  occurs at a diameter
of about $6a$ at $\beta=6.0$ which again  corresponds to  $r_0$. 


 We should like to explore also the lattice spacing dependence of  our
results since, even with an improved fermionic action, some residual 
discretisation effects are expected at $\beta=5.7$. We chose as
parameters $\beta=5.9$ with $c_{SW}=1.5$  and $\kappa_2=0.1375$ with a
$12^3 \times 24$ lattice.  Here we expect $r_0/a \approx 4.5$ which
implies that the lattice spacing is 2/3 of that at $\beta=5.7$. Thus the
spatial lattice size corresponds to  $L=8$ at 5.7. This, unfortunately,
means that the finite size effects  on the excited states will still be
significant, as found above. For this reason we do not pursue this study
here, waiting instead until we have resources  to enable us to study
larger spatial sizes than $12^3$.

\subsection{Baryons}

In addition to mesons, we are also interested in $Q L_1 L_2$ baryons
where $L_i$ refers to a $u,d$ or $s$ quark and $Q$ is a static quark.
Since $b$ quarks are close to static,  we describe such states by that
name. We only  consider states with no orbital angular momentum here, so
in the static limit these baryonic states can be described by giving the
light quark spin and parity. The  lightest such state is expected to be
the $\Lambda_b$ baryon with light quarks of $S^P=0^+$ which   can be
created by the local operator with Dirac index $i$:
 \be
   \epsilon_{abc} Q_{ia} u_{jb} (C\gamma_5)_{jk} d_{kc}
 \ee
 We treat the two light quarks  as different, even if they have the same
mass on the lattice. Experimentally, these states will be the $\Lambda_b$ 
and $\Xi_b$ for $ud$ and $qs$ light quarks respectively, where $q$ means 
$u$ or $d$.  

In the static limit for the $b$ quark, there will be only one other 
baryonic combination~\cite{stella}  with no orbital angular momentum,
namely the $\Sigma_b$ and degenerate $\Sigma_b^*$ with light quarks of
$S^P=1^+$ created by
 \be
   \epsilon_{abc} Q_{ia} u_{jb} (C\gamma_r)_{jk} d_{kc}
 \ee
 In this case we average over the three spatial components $r$.
Experimentally  these states will be the $\Sigma_b$, $\Sigma_b^*$,
$\Xi_b'$, $\Xi_b^*$,  and $\Omega_b$, $\Omega_b^*$ for $qq$, $qs$ and
$ss$ light quarks respectively.

There are some computational issues. As two light quarks are involved,
we need to  use different stochastic samples for each.  These can be 
obtained variance improved fields $v$ in $R$ and $R'$
but a little care is needed when both light quarks have the
same mass. One way to grasp the subtlety is to imagine that there are
two quarks with different flavours. Then one has to split the sum over
samples $\alpha = 1 \ldots N_s$ into subsets for each flavour. If these
subsets are independent, one obtains propagators of each flavour with no
bias. In practice if the stochastic estimators $\phi^\alpha$ are
independent of $\alpha$, one can calculate the required propagators from
 \be
G_{ji}^\alpha G_{j'i'}^\beta = \sum_{\alpha \ne \beta} 
(Q_{ik}v_k^\alpha)^* (Q_{i'k'}v_{k'}^\beta)^* v_j^\alpha v_{j'}^\beta.
\label{lambda}
 \ee
 where $k,\ k' \in R$ and $j,\ j' \in R'$.
The $\Lambda_b$ correlator is then easily constructed by multiplying
two light quark propagators from eq.~(\ref{lambda}) by the gauge
links corresponding to a heavy quark propagator with the appropriate 
$\gamma$ matrix contractions. Note that since we save the stochastic
samples and their variance reduced  fields, we need very little extra
computation to study this area. In order to be sure that there is full
independence of the $\alpha$ and $\beta$ samples we chose
$|\alpha-\beta| > 3$ with our stochastic Monte Carlo parameters
described previously. This reduces the statistics very little while
increasing the number of sweeps between samples.

\begin{figure} [t!]

\vspace{11cm}
\includegraphics{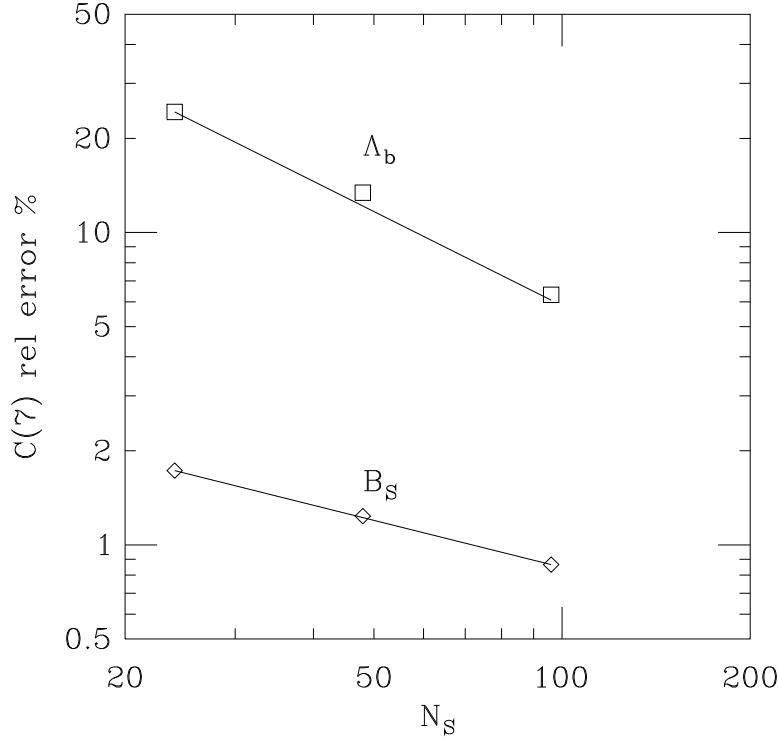}

 \caption{Relative error as a function of the stochastic sample size
$N_s$. For the S-wave B meson containing only one light quark the
decrease is consistent with $N_s^{-1/2}$, while for the $\Lambda_b$
baryon the decrease is consistent with $N_s^{-1}$. The results are from
one gauge configuration  at $\beta=5.7$ with clover fermions at
$\kappa=0.13843$ on a 12$^3$ 24 lattice for the correlation of local
hadronic operators at $t=7$. }
 \label{figure:ns}
\end{figure}

 Note that because approximately $N_s^2$ samples are used, the error on
the stochastic method  decreases faster with $N_s$ for baryons. This is
illustrated in Fig.~\ref{figure:ns} for one gauge configuration. Here 
the local operators are used for S-wave B meson and $\Lambda_b$
respectively  and the correlator from 1 gauge configuration at $t=7$
with $N_s$ stochastic samples of mass $\kappa_2$ is plotted with errors 
coming from a jackknife analysis.

For our study of the spectrum we use, as previously, $N_s=24$ from 20 
gauge configurations. For each of the light quarks we use  either a
local coupling or a sum of straight fuzzed links of length $l=1$ where 2
iterations of our fuzzing were used. This gives three hadron operators 
(neither, one and both light quarks fuzzed) which we employ at source
and sink. Using both available hopping parameters for the light quark
gives the results shown in Table~\ref{table:leff}. Note that the results
with mixed hopping parameters (labelled 12) have higher statistics since
the full set of  stochastic samples are used (i.e. $N_s^2$).  The mass
values in Table~\ref{table:leff}  come from 2 state fits to the matrix
of correlators in the $t$ range shown. We chose $t_{\rm max} < 12$ in 
the fit because the signal was too noisy at larger $t$.

 The dependence of these baryon masses on the light quark content
appears consistent  with the mesonic results in Table~\ref{table:meff}
that  the mass in lattice units is 0.037 heavier when a light quark with
$\kappa_2$ replaces one with $\kappa_1$. The exception is  that the
result for $\Sigma_b$ with mixed hopping parameters seems anomalously
light - this appears to be a statistical fluctuation caused by our
limited number of  gauge samples.

 We may also  explore the Bethe Saltpeter wavefunctions in a similar way
as for the B mesons. The additional feature for these baryons is that
there are two light quarks  and so a definition of radius is not
unambiguous. We use two types of operator, (i) with one light quark 
fuzzed by a superlink of radius $l$ and the other at 0, and (ii) with
both light quarks at radius $l$.  We then   varied the fuzzing radius
$l$ and extracted the ground state coupling.  One feature is the same as
that found for light baryons~\cite{fuzz} - namely the distributions of
the two  cases are similar if the radius in the double fuzzed case is
increased  by a factor of $\sqrt{2}$ which is a simple way to take into
account the mean squared radius of the  three dimensional double fuzzed
case. With this interpretation,  some results for the $\Lambda_b$ are
included in Fig.~\ref{figure:wf}. We  find a very similar distribution for
the $\Sigma_b$.

 We also evaluated the baryonic  correlations on spatial lattices with
$L=8$. They are similar to those found for $L=12$ but, because of  the
more limited statistics and time extent,   it is difficult to extract a
stable signal from the fit so we are  unable to quantify the finite size
effects on the mass.


\begin{table}[hbt]

\begin{center}
\caption{$\Lambda_b$ and $\Sigma_b$ effective masses.}

\begin{tabular}{llll}

 Baryon &$t$-range & $\kappa$&$am$\\
\hline
$\Lambda_b$&4-9&11& 1.435(37) \\
$\Sigma_b$&5-9&11& 1.514(52)  \\
$\Lambda_b$&4-9&12& 1.476(35) \\
$\Sigma_b$&5-9&12& 1.493(25)  \\
$\Lambda_b$&4-9&22& 1.514(31) \\
$\Sigma_b$&5-9&22& 1.621(27)  \\


\end{tabular}
\label{table:leff}
\end{center}
\end{table}

\subsection{Comparison with earlier results and experiments}

 Since we are using a quenched Wilson action at $\beta=5.7$, for
quantities  defined in terms of gauge links, there will be order  $a^2$
effects in mass ratios from this discretisation. For the $0^{++}$
glueball,  the dimensionless product with $r_0$ has been extensively
studied and substantial  order $a^2$ effects are observed~\cite{glue}.
Indeed this ratio at $\beta=5.7$ is  only 65\% of the continuum value. 
It is commonly thought that the $0^{++}$  glueball has especially large
order $a^2$ effects, so that other quantities  may well have smaller
departures at $\beta=5.7$ from their continuum values.

 Discretisation effects arising from the fermionic component are of
order  $a$ for a Wilson fermion action. A SW-clover improvement term
reduces  this and a full non-perturbative choice~\cite{NP} of the
coefficient $c_{SW}$  would remove this order $a$ discretisation error.
As discussed above,  at $\beta=5.7$, the non-perturbative improvement
scheme can not be implemented because of  exceptional configurations. A
more heuristic tadpole-improvement scheme can  give estimates in this
region and that is what we have employed here. From  a thorough study of
the light quark spectrum in  this scheme in ref.\cite{scale} with the
parameters we use here, we can establish the  region that we are
exploring. 

 Because of the significant discretisation effects in the region  of
parameters we are using, a definitive study would require results  at
larger $\beta$ so that extrapolation to the continuum limit would be 
possible. In this exploratory study, we present results at the coarse 
lattice spacing to show the power of the stochastic inversion method  in
extracting signals for hadrons. Since the study of light quark hadrons
at this lattice spacing~\cite{scale}  does show qualitatively the
features of the continuum limit, we  present our  results in way that
allows comparison with experimental data.

The extrapolation to the chiral limit  is uncertain in the quenched
approximation because of effects from  exceptional configurations and
because of possible chiral logs. Thus, as well as giving the chirally
extrapolated results,  we  present our results without extrapolation to
avoid that source of systematic error. This can be achieved by
interpolating to the strange quark  mass for the light quark. Following
ref.\cite{scale}, we define the strange quark mass by  requiring
$m_V(\bar{s}s)/m_P(\bar{s}s)=1.5$,  and assuming that the quark mass is
proportional to the squared pseudoscalar  mass, which gives  $\kappa_1$
as $0.91(2)m_s$ and $\kappa_2$ as $1.77(4)m_s$. Hence  we can extract
results for  strange light quarks by interpolation (as 90\% those  with
$\kappa_1$ and 10\% those with $\kappa_2$).

 Equating $m_V(\bar{s}s)$ to the $\phi$ meson gives the scale $a=0.82$
GeV$^{-1}$.  This can be compared with the scale obtained from using
$r_0$ (see below) which gives  $a=0.91$ GeV$^{-1}$. The scale obtained
from different observables is likely  to be different because of the
coarse lattice spacing, and indeed differences  of order 10 \% are seen
in ref.\cite{scale} when comparing with known continuum  results. 

 Another issue is the possibility of  finite size effects. The study of
the  light quark spectrum~\cite{scale} shows no sign of any significant
difference going from spatial  size $L=12$ to $L=16$. This encourages us
to expect that some of our results for $L=12$  may be close to those for
infinite spatial volume. We can check this by comparing with $L=8$  and
by looking at the Bethe-Saltpeter wavefunctions for the different
states, as discussed above.

\begin{figure}
\leavevmode

\flushleft
\epsfysize=6.5cm
\epsfbox{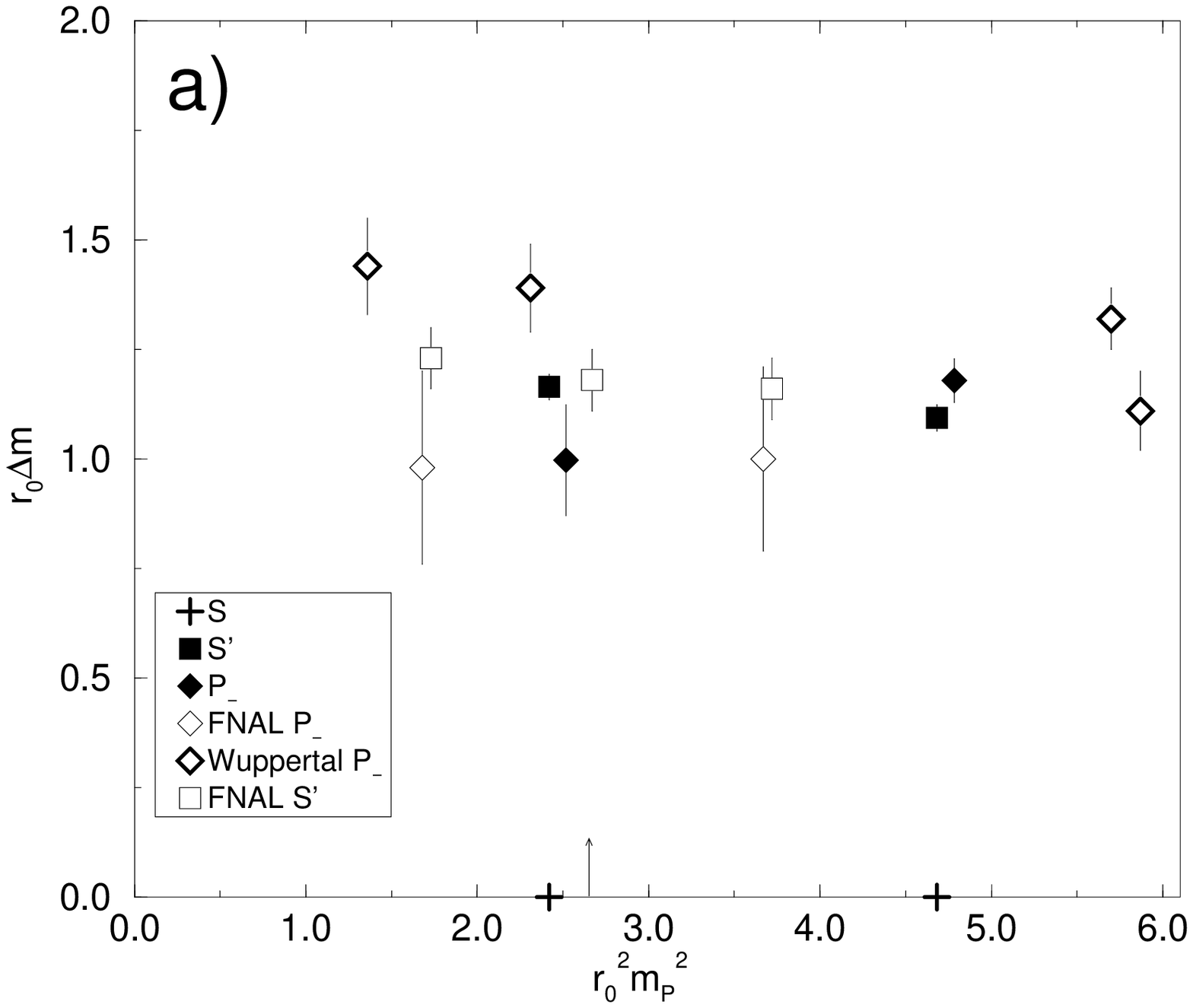}
\vspace{-6.5cm}
\flushright
\hspace{0.1cm}
\epsfysize=6.5cm
\epsfbox{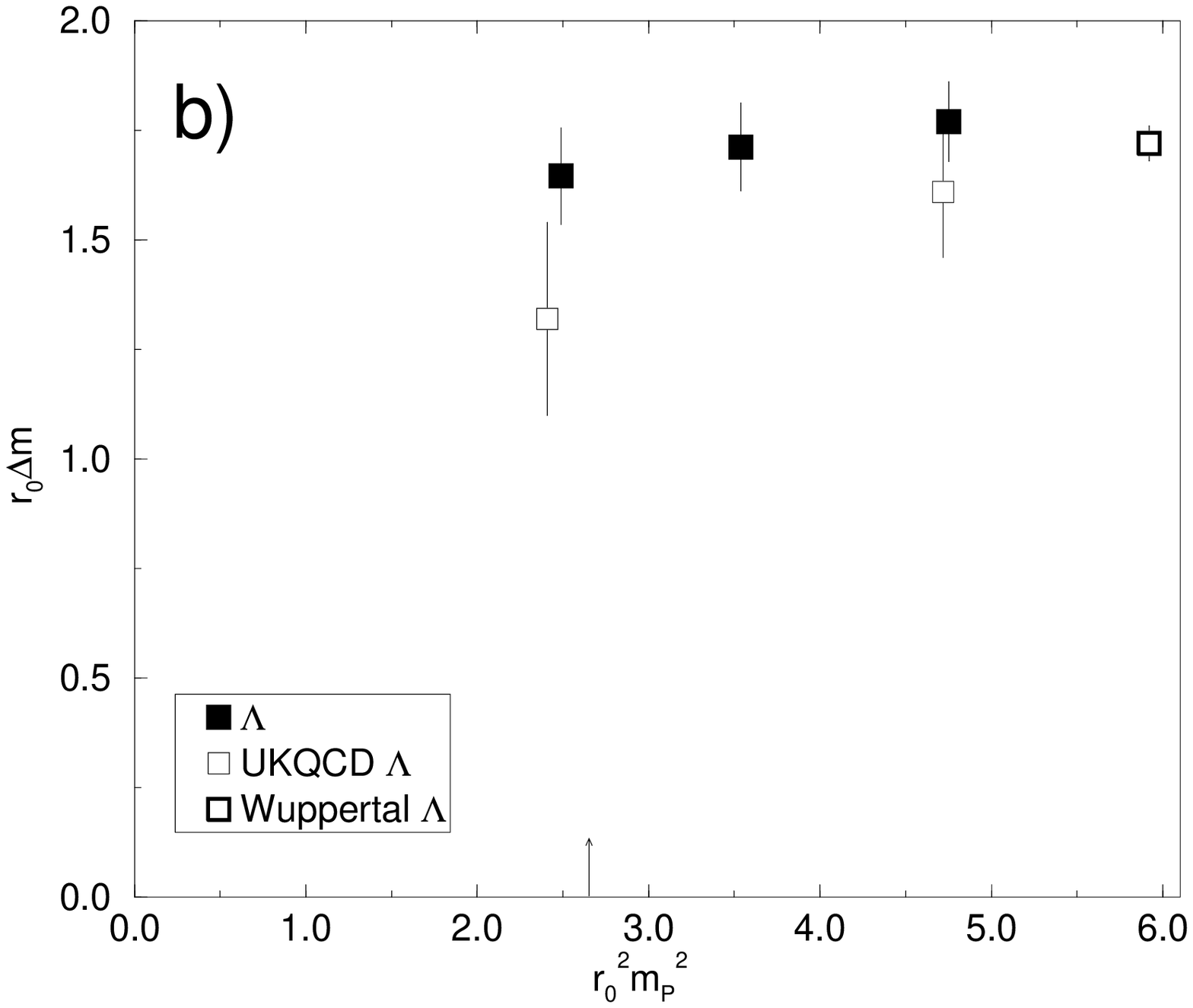}

\flushleft
\epsfysize=6.5cm
\epsfbox{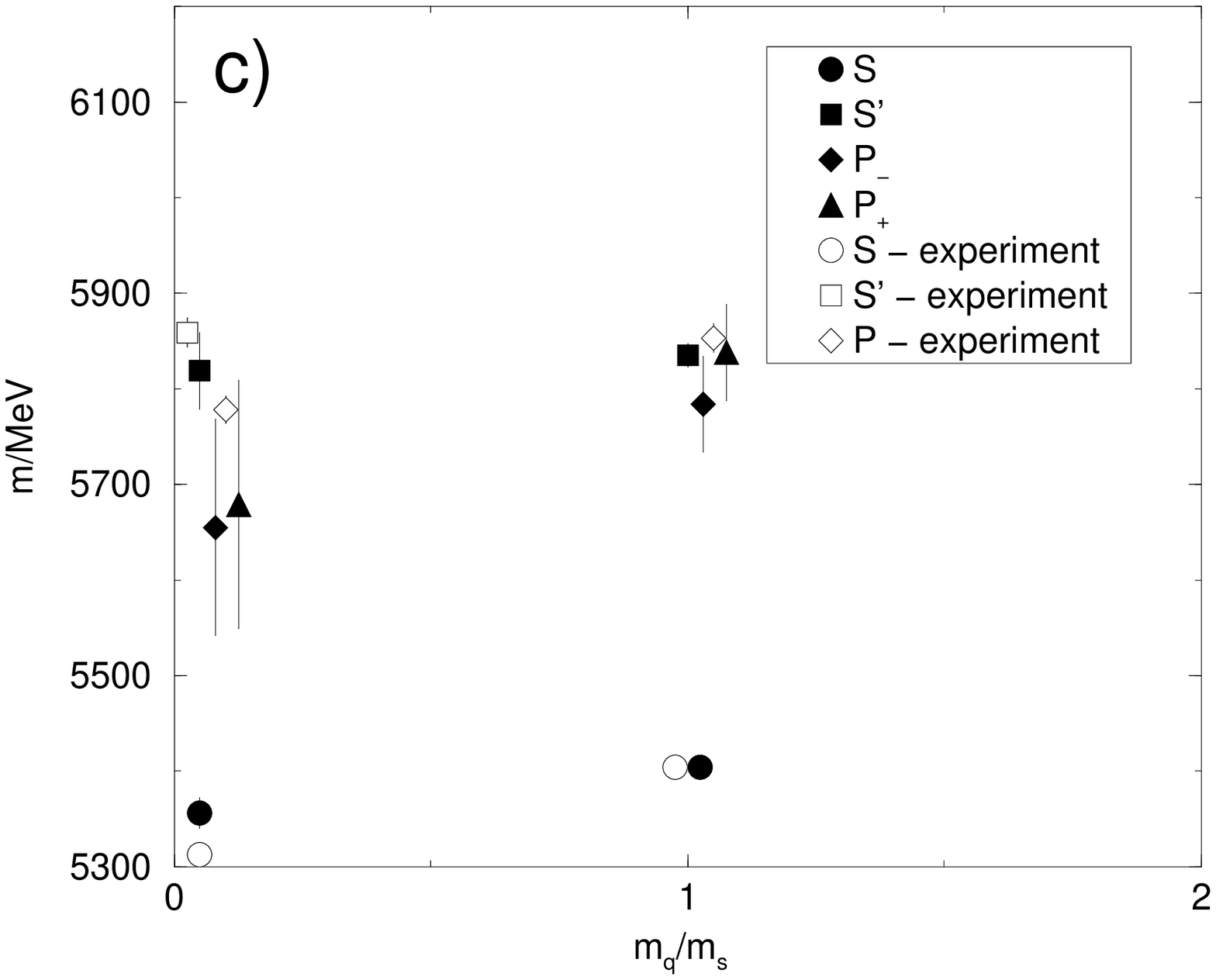}
\vspace{-6.5cm}
\flushright
\hspace{0.1cm}
\epsfysize=6.5cm
\epsfbox{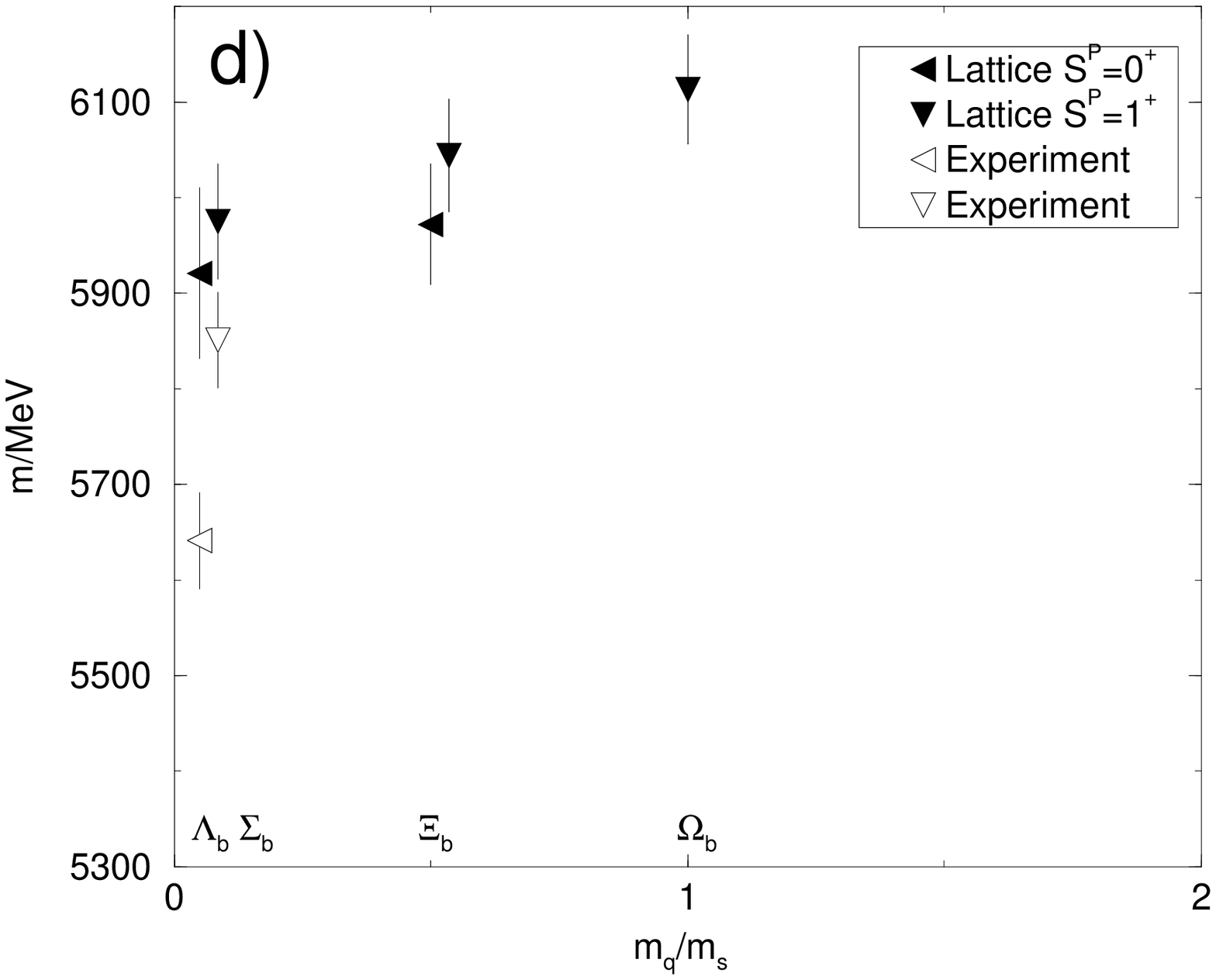}

\vspace{-0.5cm}

\caption{(a, b) Comparison of our results to earlier lattice results in
the static  limit~{\protect\cite{wuppertal,fnal,ukqcd}} where we have
plotted the mass splittings between excited states and ground state at a
given light quark mass in units of $r_0$.  The horizontal scale  is
proportional to the light quark mass (the average light quark mass for
baryons). The strange quark mass  is also shown. 
 (c, d) Comparison of  our results  to
experiment~{\protect\cite{pdg,expt,sigmab}} where the centre of gravity
of states degenerate in the static limit is plotted where available.
The  origin is set by the $B_s$, $B^*_s$ mass. The mass scale of our
lattice  results is evaluated as described in the text using
$a(5.7)=0.91$ GeV$^{-1}$.
 }
 \label{figure:mBcomp} 
 \end{figure}

We first compare our  results, using $r_0/a(5.7)=2.94$, to lattice
results obtained using usual inversion techniques. In
Figure~\ref{figure:mBcomp}(a,b) we have plotted results which several
other groups \cite{fnal,ukqcd,wuppertal} have obtained in  the static
limit  using much  more computing resources. Note that some of these 
earlier works~\cite{wuppertal,fnal} only use un-improved Wilson fermions.
Our results are clearly consistent with earlier lattice results, within
the errors quoted. However, we have  smaller error bars and are able to
obtain reliably several excited states, which is not generally true for
the earlier work in the static limit.

 We also present, in Table~\ref{table:mev} and in 
Figure~\ref{figure:mBcomp}(c,d), our results in MeV by assuming that 
the scale is set by $r_0=2.68$ GeV$^{-1}$. To avoid self energy effects
from the static quark, mass differences are evaluated and we choose as
input  the ground state S-wave B mesons with strange light quarks.  In
the heavy quark limit, this S-wave B meson should  be  identified with
the centre of gravity of the pseudoscalar and vector B mesons. 
 There will be an overall  scale error which may be significant and is
expected to be at least 20\% on energy differences. We also provide
results extrapolated to massless light quarks assuming that the B meson
masses are linear in $m_P^2$ - where in the table $s$ and $q$ refer to
strange light quarks and massless light quarks respectively. As
discussed above there will be significant systematic errors from this
chiral  extrapolation if it is not purely linear, as well as the
statistical errors  from bootstrap shown in the table. The chiral
extrapolation  for higher lying states may also be effected by finite
size effects too. A further issue is the residual effect in heavy quark
effective theory from treating the $b$ quark as static - this has been
estimated to be  around 40  MeV for the $\Lambda_b - B$ mass
difference~\cite{stella} and around 30-50 MeV for the S-P splitting in B
mesons~\cite{khan} which gives an order of magnitude  estimate of this
source of error.

\begin{table}[hbt]
\begin{center}
\caption{Heavy-light Spectrum in MeV}

\begin{tabular}{lcllll}  
State & $J^{P}$  &    latt($s$)&   expt($s$) &  latt($q$)&  expt($q$)  \\
\hline
$B(S)$  &  $0^-$, $1^-$  & input & 5404 &   5356(16)    & 5313  \\
$B(S')$ &  $0^-$, $1^-$ & 5835(12)&  - &   5819(40) &  5859(15)\\
$B(P_-)$ &  $0^+$, $1^+$ & 5784(50) & 5853(15) & 5655(113)  &    5778(14)\\
$B(P_+)$ & $1^+$, $2^+$ & 5838(50) &  5853(15)   &   5679(130)  & 5778(14) \\
$B(D_-)$ & $1^-$, $2^-$& 6001(25)&-&5934(43)&-\\
$B(D_+)$ & $2^-$, $3^-$& 6349(60)&-&6392(120)&-\\
$B(F)$ & $2^+$, $3^+$, $4^+$& 6475(50)&-& 6409(83)&-\\
$\Lambda_b$ & ${1 \over 2}^+$ & 6023(41) & - & 5921(89) & 5641(50) \\
$\Sigma_b$ & ${1 \over 2}^+$, ${3 \over 2}^+$ & 6113(57)& -& 5975(60)&
  5851(50)\\  
\end{tabular}
\label{table:mev}
\end{center}
\end{table}

 Comparing our results, remembering that only the statistical errors are
 included in Table~\ref{table:mev}, with experiment~\cite{pdg, expt,
sigmab}, we see several discrepancies. Note, however, that some
assignments of excited states  experimentally are rather uncertain, for
instance the excited strange B meson seen at 5853 MeV has no definite
$J^P$.

One feature is that the dependence of the B meson mass on the light
quark mass  is smaller than experiment. To compare with different 
lattice groups, we  evaluate the dimensionless quantity which is the
slope of $m_B$ versus  the squared light-quark pseudoscalar meson mass, 
where a common light quark is used in the heavy-light and light mesons,  
 $$
  J_b = { 1 \over r_0} {d m_B \over d m_P^2 }
 $$
 From our two $\kappa$ values, we find $J_b=  0.048(13)$. Previous
lattice determinations in the static limit~\cite{fnal,ukqcd,ape}  show
consistency with a linear dependence and give values of $J_b$ between
0.05 and 0.08. The results of  ref.\cite{fnal} with Wilson light quarks
show some evidence for an increase of $J_b$ from  $\beta=5.7$ to 6.1.
The experimental value of  the $B_s$ to $B_d$ mass difference is 90 MeV
and, using  the string tension ($\sqrt{K}=0.44$ GeV) to set the scale of
$r_0$, gives $J_b= 0.074$.  The mass difference of $D_s$ to $D_d$,
taking the centre of gravity of the vector and pseudoscalar states, is
about 103 MeV and since  the corrections to  the HQET limit are expected
 to behave like $1/m_Q$, the change from the $b$ quark to  static quarks
should be very small for this quantity - of order 2\%. In summary, our
quenched lattice result  for $J_b$ appears to be low compared to
experiment - as also found by ref.\cite{ape}.

This is reminiscent of the case for light quark mesons, where in the
quenched approximation,  a similar property is found -
summarised~\cite{J} by the $J$ parameter (proportional to slope of $m_V$
versus  $m_P^2$) which is typically 0.37 rather than the experimental
value of 0.48(2). This is equivalent to  `the strange quark problem'  of
quenched QCD where a consistent way to set the strange quark mass is not
 possible. This seems to carry over, in our work,  to the heavy-light
spectrum. As well as in the $B$ spectrum as described above,  we also
see the same effect in the light quark mass  dependence of the
$\Lambda_b$ baryon - as discussed previously.

 The spectrum of mesonic excited states can be compared with
experimental data which has been interpreted as showing evidence for 
the assignments given~\cite{expt,pdg,sigmab}. Our result for the radial
excitation appears to agree well  while the P-wave results agree
qualitatively too. We give our predictions for the higher orbital
excitations too. The finite  size effects on these excited states should
be more significant than on  the ground state and, thus, this source of 
systematic error may only  be removed by exploring even larger spatial
volumes than here.   

 Our results for $\Lambda_b$ are significantly larger than experiment.
We have checked that there do not appear to be significant systematic
errors  from extracting the ground state signal. Our results for the
baryonic  correlations from $8^3$ spatial lattices do not suggest any
very strong finite size effects either. Moreover, as seen in
Fig.~\ref{figure:mB}, we agree with other lattice determinations in
the static limit  although the statistical significance of these earlier
studies is quite low. The discrepancy in conclusion compared with
ref.\cite{ukqcd} is in the  extrapolation to the chiral limit.   The
lattice results agree within errors but the slope of the  $\Lambda_b$ mass
versus $m_P^2$ is rather different which leads to the  much  lower mass
value in the chiral limit from ref.\cite{ukqcd}.

As well as lattice results in the static limit, studies have been
undertaken with  propagating quarks. The conventional method implies a
significant extrapolation  in heavy quark mass to reach the $b$ quark -
and even more so to compare with  static quarks.  More relevant to our
work is the NRQCD method which allows heavy quarks to be used explicitly
 on a lattice.  This NRQCD method has also been used to study this  area
and has reported~\cite{khan} preliminary mass values for $b\bar{q}$  S
and P-wave mesons  and for $bqq$ baryons with light and strange quarks
in qualitative agreement with experiment.  Their results
suggest a level ordering with the $P_+$ state above the $P_-$ by about
200 MeV with a significance of $4\sigma$.  Their results for the
baryonic levels are lower than ours, in better agreement with experiment.

\section{Non-static systems}

The simplest such situation is the  spectrum of mesons made from 2 light
quarks.  To test our approach, we have compared the results from
stochastic maximal variance reduction with conventional
methods~\cite{scale}. 

As a cross check, we have measured pseudoscalar and vector meson masses 
for clover fermions on 12$^3$ 24 lattices at the same parameters as 
ref.\cite{scale}. We use superlinks of length $l=3$ made from links 
with 5 fuzzing iterations as well as local observables and  a two state
fit with the excited state  fixed at $am=1.75$ to stabilise the fit. We
obtain from $N_s=24$ samples on 20 gauge configurations the results
shown in Table~\ref{table:PV}. The  values from conventional inversions
on 500 gauge configurations~\cite{scale} clearly have much smaller
errors  than our method using 20 gauge configurations. By increasing our
 sample size $N_s$, we could improve our signal/CPU, since the noise
decreases as $1/N_s$, but the conventional  method works very well here.
 In principle, one gains by increasing $N_s$ only  until the noise from
finite sample size is comparably small to  the inherent fluctuations 
from gauge configuration to gauge configuration. This limit on $N_s$
will depend on the observable  under study and other parameters and,
after exploring values of $N_s$ up to 96,  we find that in general it is
greater than 96.

The main reason that the stochastic  maximal variance reduction is so
noisy is that our variance reduction  is in terms of the number of links
(in the time direction) from the source planes.  The zero-momentum meson
correlator involves a sum over the whole of the source and sink
time-planes.  The noise from each term in this  sum  is similar even
though the signal  is small at large spatial separation differences.
Thus we get noise from  the whole spatial volume whereas the signal is
predominantly from a part of the volume.  This same problem also plagues
large spatial volume studies of glueballs  and the solution~\cite{mt} in
that case is to evaluate the non-zero momentum correlators  since they
are related to the well measured part at relatively small spatial
separation. This approach should help equally with our stochastic
maximal variance reduction method.

\begin{table}[hbt]

\begin{center}
\caption{Pseudoscalar and Vector meson masses}

\begin{tabular}{lllll}

Meson &$\kappa$& $t$-range& $am$   & ref.\cite{scale}\\
\hline
$P$&11 &3-12& 0.523(30) &  0.529(2)\\
$V$&11 &3-12& 0.731(87)   & 0.815(5)\\
$P$&22 &3-12& 0.740(36)   & 0.736(2)\\ 
$V$&22 &3-12& 0.977(48)   & 0.938(3)\\ 

\end{tabular}
\label{table:PV}
\end{center}
\end{table}

 Even though the meson mass spectrum is rather noisy compared to 
conventional inversions, there is a substantial  gain from using 
our all-to-all techniques when exploring matrix elements of mesons. In 
this case three or more light quark propagators will be needed and 
they must be from more than one source point. This is straightforward to
evaluate using our stochastic techniques with our stored sample fields.
We intend to explore this area more completely elsewhere.

One of the  problems, in the quenched approximation, with the
conventional approach to light quark propagators  is that exceptional
configurations cause huge fluctuations  in the correlation of hadrons,
especially of pions, at hopping parameters close to the chiral limit.
These exceptional configuration problems are associated with regions of 
non-zero topological charge.   Using all-to-all propagators may smoothen
these fluctuations  somewhat  in that the average over the spatial
volume will fluctuate less than the  propagator from one site.
Eventually, however, this problem of exceptional configurations can only
be solved unambiguously  by using dynamical quark configurations.

\section{Conclusions}

We have established a method to study hadronic correlators using 
stochastic propagators which can be evaluated from nearly all sources to
nearly all sinks and  which allow the correlations to be obtained with
relative errors which do not  increase too much at large time
separation.
 In this exploratory study, we have considered light quark propagators
from about 1 million sources for each $\kappa$ value.  The amount of
resources we have used is minimal. The total CPU time is roughly 10
Mflops-year, and the total disk space needed for all our results is 17
Gbytes.

 We find that for hadrons involving one static quark, our approach is 
very promising. We have been able to explore the spectrum of excited B
mesons and heavy quark baryons in  detail, albeit at a rather coarse
lattice spacing. The results go beyond previous lattice work, in particular 
in exploring higher orbital angular momentum excitations. We find evidence 
for a linear dependence of mass on orbital angular momentum for 
heavy-light mesons up to F-waves. 

For the light quark mass dependence we  find that the slopes of the
heavy-light meson and baryon masses versus  the squared pseudoscalar mass
with that light quark are both  significantly less than experiment. A
similar feature has been found  for light-light vector mesons and
baryons with a similar reduction of slope of about 70\%. A common 
explanation would be  that the  quenched approximation is mainly
deficient in providing light-light  pseudoscalar masses. This is not
unreasonable  since both the effect of disconnected diagrams (the $\eta$
splitting from the $\pi$) and the effect of exceptional configurations
are expected to be most important for pseudoscalar mesons.

We determine the heavy-light baryon to meson mass differences which we
find  to be significantly larger than experiment. This may arise from
discretisation effects at  our rather coarse lattice spacing, a
non-linear extrapolation to the chiral limit  or enhanced finite size
effects. Note that for light baryons, recent precise data~\cite{cppax} 
show significant non-linearity for the $J^P={1 \over 2}^+$ states which
has the effect of reducing the lattice mass prediction  in the chiral
limit compared to a linear extrapolation. This effect, if present  for
heavy-light baryons, would go some way to explain our discrepancy.

To establish our results more fully, we need to study the approach to
the continuum limit and  to check on finite size effects.  An increase 
in the number of gauge configurations would also  allow a more thorough
analysis of errors. Thus we would need to explore larger  lattices at
smaller lattice spacing. This is straightforward in principle,  but
involves a non-trivial re-organisation of the logistics of creating  and
storing the stochastic samples.

 The approach can easily be extended to other cases involving static
quarks -  particularly matrix elements and interaction energies between
two  B mesons.  Another application is to study the bound states  of a
static adjoint source with light quarks.

 One motivation for this work is that dynamical fermion configurations 
are very expensive computationally to create. Thus one should use  fully
the information contained in the gauge configurations available. Our
method works  straightforwardly with such dynamical fermion gauge
configurations  - thus it is the method of choice to explore these
configurations most fully by evaluating correlations from all sources.

 One potential advantage of stochastic methods to determine  propagators
is that disconnected diagrams are accessible. Unfortunately, as we  have
explained, our maximal variance reduction technique does not help to
reduce the noise of any component of the correlation which has a fermion
loop  with common sink and source. This area of research will need other
variance reduction techniques than those we have presented here.

\section{Acknowledgements}

We acknowledge support from PPARC grant GR/K/53475 for computing time at
DRAL. JP wishes to acknowledge support from the Magnus Ehrnrooth
Foundation, the Alfred Kordelin Foundation, the V\"ais\"al\"a Foundation
and the H. C. Baggs Fellowship.

\appendix

\section{Appendix: Construction of $B$ meson operators in the static limit.}

  In the heavy quark limit, the $\bar{Q} q$ meson which we refer to as a
`B' meson, will be the `hydrogen  atom' of QCD. Since the meson is
made from non-identical quarks, charge conjugation  is not a good
quantum number.  States can be  labelled by  $L_{\pm}$ where the
coupling of the light quark spin to the orbital angular momentum  gives 
$j=L\pm {1 \over 2}$. In the heavy quark  limit these states will be
doubly degenerate since the heavy quark spin interaction can be 
neglected, so  the  $P_{-}$ state will have  $J^{P}=0^+,\ 1^+$ while
$P_{+}$ has $J^{P}=1^+,\ 2^+$, etc.

  We now describe lattice operators to construct these states. For the
generic construction, following the conventions of \cite{lm}, we use
nonlocal operators for the B meson and its excited states. This will
enable us to  study also the orbitally excited mesons. The operator $B$
we use to create such a $\bar{Q} q$ meson on the lattice is defined on a
timeslice $t$ as  
 \be B_t = \sum_{x_1, x_2} \bar{Q}_({\bf x}_2,t) 
    P_t({\bf x}_1, {\bf x}_2) 
    \Gamma 
   \, q({\bf x}_1,t).
 \ee
 $Q$ and $q$ are the heavy and light quark fields respectively, the sums
are over all space at a given time $t$, $P_t$ is a linear combination of
products of gauge links $U$ at time $t$ along paths $P$ from ${\bf x}_1$
to ${\bf x}_2$, $\Gamma$  defines the spin structure of the operator.
The Dirac  spin indices and the  colour indices are implicit.

 In this work we choose paths $P_t$ which are specific combinations of  a
product  of fuzzed links in a straight line of length $l$. The
appropriate symmetry  for the cubic rotations on a lattice with a state
of zero momentum are  given by the representations of $O_h$.  The
relationship of these representations to those of SU(2) can be  derived
by restricting the SU(2) representations to the rotations  allowed by
cubic symmetry and classifying them under $O_h$. This  process (called
subducing) yields the results (tabulated to $L$=4):

\begin{tabular}{cc}
$L$=0 &$ A_1$ \\
$L$=1 &$ T_1$ \\
$L$=2 &$ E \ T_2$ \\
$L$=3 &$ A_2 \  T_1\  T_2$ \\
$L$=4 &$ A_1 \  E \ T_1 \  T_2$ \\
\end{tabular}

 So that an $L=3$ excitation can be extracted by looking at the $A_2$ 
representation, for example. 

For our lattice construction, we define the sum and difference of the
two such paths in direction $i$ as $s_i$ and $p_i$  respectively (the
latter is in the $T_1$ representation). The  combinations appropriate
for the discrete group of cubic rotations are then the $A_{1}$ symmetric
sum  $S=s_1+s_2+s_3$ and the $E$ combinations of $a_i$ which can be
taken as $E(a_i)=a_1-a_2$  and $(2 a_3-a_1-a_2)/\sqrt{3}$.

 The appropriate operators for B mesons in the static limit are then
 
  $$ S:\ \ \ \bar{Q} \gamma_5 S q\  \ \hbox{or}\ \ \bar{Q} \gamma_i S q
  $$
  $$P_{-}:\ \ \ \bar{Q} 1 q\  \ \hbox{or}\ \ \bar{Q} \sum_i\gamma_i p_i q
  $$
  $$P_{+}:\ \ \  \bar{Q} E(\gamma_i p_i) q
  $$
 with no sum on $i$.
  $$ D_{\pm}:\ \ \ \bar{Q} \gamma_5 E(s_i) q
  $$
 Note that this operator is a mixture of both $D_{\pm}$ states.

 In order to access higher spin states, following~\cite{lm}, we also
consider L-shaped paths $P_t$ where each side of the L has the same
length. We take linear  combinations of these in the $T_2$
representation (paths $t_i$ where $i$ is direction of normal to plane of
paths) and in the $A_2$ representation (paths $a$). This allows  
us to separate the $D_{\pm}$ states since

$$ D_{-}:\ \ \ \bar{Q} E(\gamma_i t_i) q
 $$
 with no sum on $i$.
$$ D_{+}:\ \ \ \bar{Q} \sum_i(\gamma_i t_i) q
 $$
 Also
$$ F_{\pm}:\ \ \ \bar{Q} \gamma_5 a q
 $$
 where $a$ refers to the sum with alternating sign of  paths to  8
corners of a spatial cube from the centre. The paths to each corner are
the sum of  the 6 routes of shortest length along the axes, combined by
projecting to  the $SU(3)$ group after addition. This gives an $L=3,\
6,..$ state. This operator is a mixture of both $F_{\pm}$ states.

 In each case we use two different fuzzing/length choices to build up
the operators. For $S$ and   $P_-$ we also have a local operator
available. We also explored  additional operators with $\sum_i \gamma_i
p_i$ factors but they do not  add anything very useful in practice. We
measure the correlations between  each of the 2 (or 3) operators at
sink and source so obtaining a matrix  of correlations which can be used
to separate the excited states from  the ground state of that quantum
number.

For off-diagonal elements there is one further subtlety. The alternate 
light quark stochastic expression introduces  extra $\gamma_5$ factors.
For the $P_-$ correlation between 1 and $\gamma_i p_i$, this will
introduce  a relative sign change as well as changing $H_- \to H_+$.
Interchanging the operators between source and sink now involves taking
$-\gamma_4 1 \gamma_4$ instead  of 1  which  introduces a
further  minus signs in this correlation.

\end{document}